\documentclass[]{aastex62}
\usepackage[]{hyperref}
\usepackage{float}
\usepackage{natbib}

\usepackage{amsmath}
\usepackage{graphicx}

\newcommand{\fluxunits}{erg cm$^{-2}$ s$^{-1}$}
\newcommand{\feii}{Fe~\textsc{II}~$\lambda2814.45$}
\newcommand{\prim}{$^{\prime}$}
\newcommand{\intunits}{erg cm$^{-2}$ s$^{-1}$ sr$^{-1}$ \AA$^{-1}$}

\shorttitle{NUV Flare Spectra}
\shortauthors{Kowalski et al.\ }

\begin{document}

\title{Spectral Evidence for Heating at Large Column Mass in Umbral Solar Flare Kernels I:  IRIS NUV Spectra of the X1 Solar Flare of 2014 Oct 25}

\author{Adam F. Kowalski}
\affil{Department of Astrophysical and Planetary Sciences, University of Colorado Boulder, 2000 Colorado Ave, Boulder, CO 80305, USA.}
\affil{National Solar Observatory, University of Colorado Boulder, 3665 Discovery Drive, Boulder, CO 80303, USA.}
\affil{Laboratory for Atmospheric and Space Physics, University of Colorado Boulder, 3665 Discovery Drive, Boulder, CO 80303, USA.}
\email{adam.f.kowalski@colorado.edu}
\author{Elizabeth Butler}
\affil{Department of Astrophysical and Planetary Sciences, University of Colorado Boulder, 2000 Colorado Ave, Boulder, CO 80305, USA.}
\author{Adrian N. Daw}
\affil{NASA Goddard Space Flight Center, Heliophysics Sciences Division, Code 671, 8800 Greenbelt Rd., Greenbelt, MD 20771, USA.}
\author{Lyndsay Fletcher}
\affil{SUPA School of Physics \& Astronomy, University of Glasgow, Glasgow, G12 8QQ, UK.}
\affil{Rosseland Centre for Solar Physics, University of Oslo, P.O.Box 1029 Blindern, NO-0315 Oslo,  Norway.}
\author{Joel C. Allred}
\affil{NASA Goddard Space Flight Center, Heliophysics Sciences Division, Code 671, 8800 Greenbelt Rd., Greenbelt, MD 20771, USA.}
\author{Bart De Pontieu}
\affil{Lockheed Martin Solar and Astrophysics Laboratory, Org. A021S, Bldg. 252, 3251 Hanover St., Palo Alto, CA 94304, USA.}
\affil{Rosseland Centre for Solar Physics, University of Oslo, P.O. Box 1029 Blindern, NO-0315 Oslo, Norway.}
\affil{Institute of Theoretical Astrophysics, University of Oslo, P.O. Box 1029, Blindern, Oslo, Norway.}
\author{Graham S. Kerr}
\affil{NASA Goddard Space Flight Center, Heliophysics Sciences Division, Code 671, 8800 Greenbelt Rd., Greenbelt, MD 20771, USA.}
\author{Gianna Cauzzi}
\affil{National Solar Observatory, University of Colorado Boulder, 3665 Discovery Drive, Boulder, CO 80303, USA.}

\section*{Abstract}
The GOES X1 flare SOL2014-10-25T17:08:00 was a three-ribbon solar flare observed with IRIS in the near and far ultraviolet.  One of the flare ribbons crossed a sunspot umbra, producing a dramatic, $\sim1000$\% increase in the near-ultraviolet (NUV) continuum radiation. We comprehensively analyze the ultraviolet spectral data of the umbral flare brightenings, which provide new challenges for radiative-hydrodynamic modeling of the chromospheric velocity field and the white-light continuum radiation.  The emission line profiles in the umbral flare brightenings exhibit redshifts and profile asymmetries, but these are significantly smaller than in another, well-studied X-class solar flare.  We present a ratio of the NUV continuum intensity to the \feii\ intensity.  This continuum-to-line ratio is a new spectral diagnostic of significant heating at high column mass (log $m/$[g cm$^{-2}] >-2$)  during solar flares because the continuum and emission line radiation originate from relatively similar temperatures but  moderately different optical depths.   The full spectral readout of these IRIS data also allow for a comprehensive survey of the flaring NUV landscape: in addition to many lines of Fe II and Cr II, we identify a new solar flare emission line, He I $\lambda2829.91$ (as previously identified in laboratory and early-type stellar spectra).  The Fermi/GBM hard X-ray data provide inputs to radiative-hydrodynamic models (which will be presented in Paper II) in order to better understand the large continuum-to-line ratios, the origin of the white-light continuum radiation, and the role of electron beam heating in the low atmosphere.  

\section{Introduction}

Flares result from a sudden magnetic reconfiguration in the atmosphere of a star, producing electrons and protons that stream  at near the speed of light along the directions of the reconfigured magnetic fields.  The flare is the burst of electromagnetic radiation and is thought to be the result of the impact of the (mildly) relativistic electrons (``beams'') with the lower, dense stellar atmosphere; the protons and ions produce nonthermal gamma ray emission \citep{Murphy1997, Hurford2006}, but their relative contributions to the multi-thermal response has not yet been established.
The pan-chromatic continuum radiation, excluding the X-rays, extreme-ultraviolet, gamma rays, and radio emission, is collectively known as the white-light continuum radiation because it appears in broadband\footnote{Various loose definitions of a ``white-light flare'' exist, including a flare that could be detected by the eye (which is broadband). In dMe stars, $U$-band flares are certainly considered white-light flares even though our eye is not sensitive to these wavelengths, and flares detected in a narrow passband of SDO/HMI are often called white-light flares, assuming that continuum radiation is the source of the HMI increase. Generally, a white-light flare is a flare that produces a change in continuum radiation that could be detected in the Johnson $U$ and/or $V$ bands.}. The white-light is typically one of the most impulsive signatures in a solar flare \citep[e.g.,][]{Hudson2006, Fletcher2007, Namekata2017, Watanabe2017} and is spatially and temporally correlated with the hard X-rays from the nonthermal electrons on the Sun \citep{Kane1985, Hudson1992, Martinez2012}.  

A common assumption is that the white-light originates from increased photospheric (or upper photospheric) radiation. However, chromospheric condensations \citep{Livshits1981, Fisher1989, Kowalski2015} with low continuum optical depth also produce broadband continuum radiation with a large jump in flux near the hydrogen Balmer limit \citep{Gan1992, Kowalski2017A}.  The capability to test model predictions of the Balmer jump in solar flares has largely disappeared, which is unfortunate because spectra of solar flares in the 1980s (primarily from the Universal Spectrograph) exhibit a variety of characteristics in the Balmer jump spectral region \citep{Hiei1982, Acampa1982, Neidig1983, Donati1984,Donati1985, Boyer1985, Kowalski2015HSG, Ondrej2017}, and we now have methods to model opacity from blended lines and dissolved levels \citep{Uitenbroek2001, Kowalski2015}.  Therefore, we must employ other spectral diagnostics that critically test these models and constrain whether the white-light results from photospheric heating, as suggested by a recent off-limb measurement of the emission height \citep[][but see \citet{Battaglia2011} and \citet{Krucker2015}]{Martinez2012}, primarily from chromospheric heating, or significant heating throughout several layers of the lower atmosphere \citep{Neidig1993B, NeidigLimb, Kleint2016}.  Direct measurements of the white-light emission heights in limb flares may be partially occulted (H. Hudson, priv.\ communication), and these events are generally difficult to compare to 1D plane-parallel flare models \citep[but see][]{Heinzel2017}.
 The specific continuum intensity over a narrow wavelength range has also been used to test electron beam heating in the lower atmosphere \citep{Heinzel2014, Kowalski2017A}.  However, the interpretation of the measured intensity can be degenerate since an optically thin continuum source with $T\sim 10^4$ K produces a radiation temperature of $4000-6000$ K in the optical \citep{KA18} and nearly 7000 K in the near-UV \citep{Kleint2016}.  Furthermore, the white-light sources may be unresolved even with current high-spatial resolution capabilities \citep{Krucker2011, Sharykin2014, Kowalski2015HSG}.

Is the photosphere heated to produce an observable amount of continuum radiation that overpowers the chromospheric flare radiation?  Evidence for very deep heating exists in dMe flares from the Balmer continuum radiation that is produced in absorption \citep{Kowalski2013} or in emission\footnote{The terminology ``in emission'' or ``in absorption'' means that the Balmer continuum flux is, respectively, greater than or less than a blackbody curve extrapolation from a fit to continuum windows at $\lambda > 4000$ \AA.} with a relatively small jump in continuum flux in the Balmer jump spectral region \citep{HawleyPettersen1991, Hawley1992, Kowalski2013, Kowalski2016, Kowalski2018}.  The mass column density\footnote{Often used to indicate the atmospheric depth.} (hereafter, ``column mass'') that is heated to $T\sim10,000$ K to reproduce these spectral properties is far smaller than the dMe photospheric column mass of log $m$/[g cm$^{-2}$] $ \sim1$ \citep{CramWoods1982, Kowalski2017B}.  
  In solar and dMe electron beam simulations that produce a flare chromosphere that is optically thin to Balmer and Paschen continuum radiation, radiative backwarming heats the (upper) photosphere but the temperature increase in these deep layers is only $\Delta T \sim 500-1000$ K \citep{Allred2005, Allred2006, Cheng2010}, and is not enough to explain the 10,000 K blackbody-like radiation in dMe flares \citep[see also][]{Kowalski2018CS20}. Furthermore, it takes some time ($\Delta t \gtrsim10$~s;  appendix of \citet{Kowalski2017A}) for the solar photosphere to heat up in these models, and it is not clear if beam heating persists in a given flare loop for such long timescales  \citep{Aschwanden1998, Nishizuka2009, Penn2016}. 
Nonthermal electrons exhibit a power-law with most electrons at $E\sim20$ keV;  if produced in the corona \citep[as implied by hard X-ray, energy-dependent timing differences;][]{Aschwanden1995ToF, Aschwanden1995ToF2, Aschwanden1996, Aschwaden1996Masuda, Aschwanden1996ToF,Aschwanden1996ToF2, Aschwanden1998ToF}, most of the electron beam energy is expected to be lost in the mid-to-upper chromosphere \citep[log $m \sim -3$ to $-5$;][]{Vernazza1981}.  If electron beam heating is shown insufficient to explain impulsive, large temperature increases in the photosphere or low chromosphere,  
then there must be additional, important sources of heating in flares, such as Alfv\'{e}n waves \citep{Russell2013, Reep2016, Kerr2016, Reep2018} or proton beams \citep{Zharkova2007, Ondrej2018}.   Instead of Balmer jump ratio measurements, we must consider other  spectral signatures that indicate significant heating at large column mass in solar flares.  Additional heating mechanisms at large column mass would be transformative improvements to the standard solar flare model and would have important implications for the heating sources in superflares in other stars and the young Sun \citep[e.g.,][]{Maehara2012, Osten2016}.  
 
To determine if low-energy electron beams are sufficient to explain the heating in the low atmosphere, we present a new measurement that can be obtained from spectra with limited wavelength coverage, as is often the case with solar imaging spectrometers.  This new measurement is the ratio of the NUV continuum to the Fe II line intensity.  The NUV continuum-to-line ratio will be compared directly to radiative-hydrodynamic modeling in Paper II of this series to infer the largest (deepest) column mass that is heated by energetic electrons (or other energy transport mechanisms) in a flare.   The flare we study is the X1 flare on 2014 Oct 25, which produced ribbons that crossed the slit of the Interface Region Imaging Spectrograph (IRIS) during a hard X-ray event, which occurred during the decay phase of the GOES X1 event.

The paper is outlined as follows.  In Section \ref{sec:data} we describe the IRIS and the Fermi/GBM data for the X1 flare. In Section \ref{sec:irisanalysis}, we describe the properties of the NUV continuum enhancements, the \feii\ line profiles, and the ratios between the NUV continuum intensity and \feii\ line-integrated intensity.  In Section \ref{sec:otherstuff}, we describe other characteristics of this flare that are important for radiative-hydrodynamic modeling: a line  of He I that has not yet been seen in solar flare spectra and the nonthermal electron power and spectral index inferred from hard X-rays.
 In Section \ref{sec:discussion}, we summarize our findings and compare the NUV spectral properties to the modeling and IRIS observations  of another well-studied X-class flare.  In Section \ref{sec:conclusions}, we conclude with several general implications from these unique solar flare spectra. Appendix A describes the continuum-to-line ratio dependence on the temperature and density in optically thin slab model approximations, and appendix B contains identifications of observed flare lines in the IRIS NUV and discussion of the observational constraints they provide for models.

\section{Data} \label{sec:data}
The 2014-Oct-25 X1 solar flare  was observed with a unique dataset that included a custom IRIS observing mode, which is described in Section \ref{sec:irisdata}. Hard X-ray observations were provided by Fermi/GBM throughout the entire flare as well (Section \ref{sec:fermigbm}).   This comprehensive dataset allows us to test RHD modeling of electron beam heating in new ways (in Paper II).

\subsection{IRIS Ultraviolet Data} \label{sec:irisdata}
The Interface Region Imaging Spectrograph \citep[IRIS;][]{DePontieu2014} observed
the GOES X1 flare SOL2014-10-25T17:08:00 ($\mu=0.85$, $x=408$\arcsec, $y=-318$\arcsec) from National Oceanic and Atmospheric (NOAA) Active Region (AR) 12192. The X-class flares from AR 12192 have been studied extensively due to the absence of associated
coronal mass ejections \citep{Thalmann2015, Inoue2016, Bamba2017, Amari2018}.  The IRIS spectra of the 2014-Oct-25 X1 flare were obtained in a  ``sit-and-stare'' mode with an exposure time of 4~s, a temporal cadence of 5.4~s, and the slit oriented in the E-W direction.  The slit jaw images (SJI) in the Mg II $h$ wing (hereafter, SJI 2832), C II, and Mg II were obtained with a cadence of 16 s.  All spectra and slit jaw images were calibrated from DN s$^{-1}$ to specific intensity (hereafter, just ``intensity'') following \citet{Kleint2016} and \citet{Kowalski2017A} by using the time-dependent effective area curves \citep{Wusler2018}.  By comparing the fiducial marks in the FUV and NUV data, we found that a (downward) shift by one pixel in the FUV spectra was necessary to align them with the NUV spectra. We found that a shift of $-0.16 $\AA\ was required to calibrate the velocity (using the Ni 2799.474 \AA\ line, as suggested by IRIS Technical Note 26) in the level 2 data that we used for our analysis. We note that more recently calibrated data does not require such a shift, since the whole IRIS archive has been reprocessed in 2017/2018 including an improved wavelength calibration. In general, it is always best to check the calibration of the wavelength using the instructions provided in IRIS Technical Note 26.

Usually, to strike the best balance between cadence and data rate, only narrow wavelength regions around specific lines of interest are read out from IRIS.  Instead, we obtained the observations with a custom observing mode that 
downlinked the full spectral range of IRIS with two-pixel binning in the dispersion direction ($\lambda = 1331.69-1358.04$ \AA\ and $\lambda=1389.52-1406.41$ \AA\ at 0.0256 \AA\ pixel$^{-1}$ in the FUV and $\lambda=2783.93-2834.95$ \AA\ at 0.051 \AA\ pixel$^{-1}$ in the NUV) and binning by two in the spatial direction (thus, 0.33\arcsec\ binned-pixel$^{-1}$ corresponding to $\sim$250 km binned-pixel$^{-1}$ at the distance of the Sun).  The full IRIS spectral range was used for a robust identification and characterization of the FUV and NUV continua in flares;  with limited wavelength ranges of typical IRIS readout modes (``linelists''), it is not always possible to accurately determine the continuum intensity due to multiple
broad, asymmetric flares lines of Mg II, Fe II and other species.
The NUV spectral range of IRIS (hereafter, IRIS NUV) has a relatively line-free continuum region at $\lambda \sim 2826$ \AA\ that has been useful as a proxy of the Balmer continuum component of the white-light radiation for comparisons to the predictions of electron beam heating models \citep{Heinzel2014, Kleint2016, Kowalski2017A}.  In this paper, we calculate this quantity and compare to the continuum level over the entire spectral range of IRIS.

\subsection{Fermi/GBM X-ray Data} \label{sec:fermigbm}
The Fermi/Gamma-Ray Burst Monitor \citep[GBM;][]{Meegan2009} provides high-time resolution (4s, or 1s in burst mode) X-ray data of solar flares from 8 keV to 40 MeV. RHESSI observed the main peak of the flare, but the imagery indicates that the loop-top source dominated the 25-50 keV emission, and RHESSI also suffered from significant pileup \citep{Kleint2017}.  We use the Fermi/GBM data late in the flare when pileup was not significant in several of the least sunward facing  detectors.

The Fermi/GBM has two types of scintillation detectors:  twelve NaI detectors that are sensitive at $E\lesssim 200$ keV and two bismuth germanate (BGO) detectors that are sensitive at $E\gtrsim200$ keV.   For spectral analysis (Section \ref{sec:thicktarg}), we use data from detector NaI\_n0 (hereafter, ``n0''), which is the third most sunward-facing detector and does not suffer from pile-up effects\footnote{NaI\_n05 is the most sunward facing detector and suffers from obvious pileup at the peak of the hard X-ray event.} over the time of the decay phase of this flare.  For light curve analysis, we also use data from detector NaI\_n4 and NaI\_n5 (hereafter, ``n4'' and ``n5'', respectively), which are more sunward-facing in the early rise phase of the flare.  The Solar Soft (SSW) IDL OSPEX  is used to retrieve and analyze the Fermi/GBM data.  First, we retrieve a detector response (rsp2) file for the time interval of 16:35-17:36.  The background is chosen before and after the X1 event (during Fermi night) in the time-intervals of 16:15-16:35 and 17:50 - 18:05.  Since the detector angles relative to the Sun change over the first few minutes of this flare\footnote{The NaI\_n0 detector changes from the least sunward facing orientation to the third most sunward facing orientation from 17:01 to 17:05, which affects the rise phase of the hard X-rays.}, we divide the OSPEX count rate by the cosine of the angle between the detector normal and the Sun. 

\subsection{Overview of the Flare}
\begin{figure}[h!]
\centering
\includegraphics[scale=0.44]{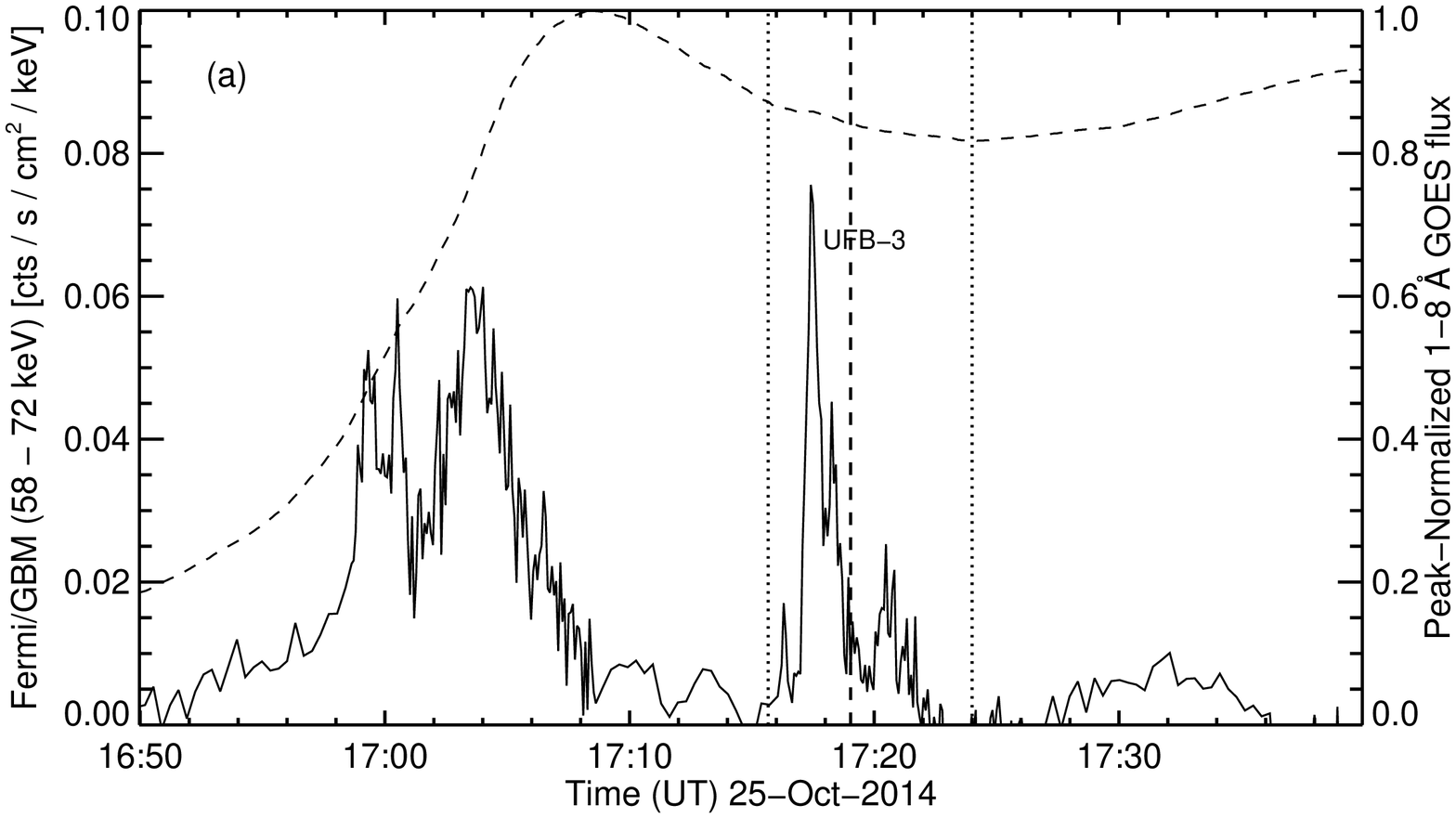}
\includegraphics[scale=0.44]{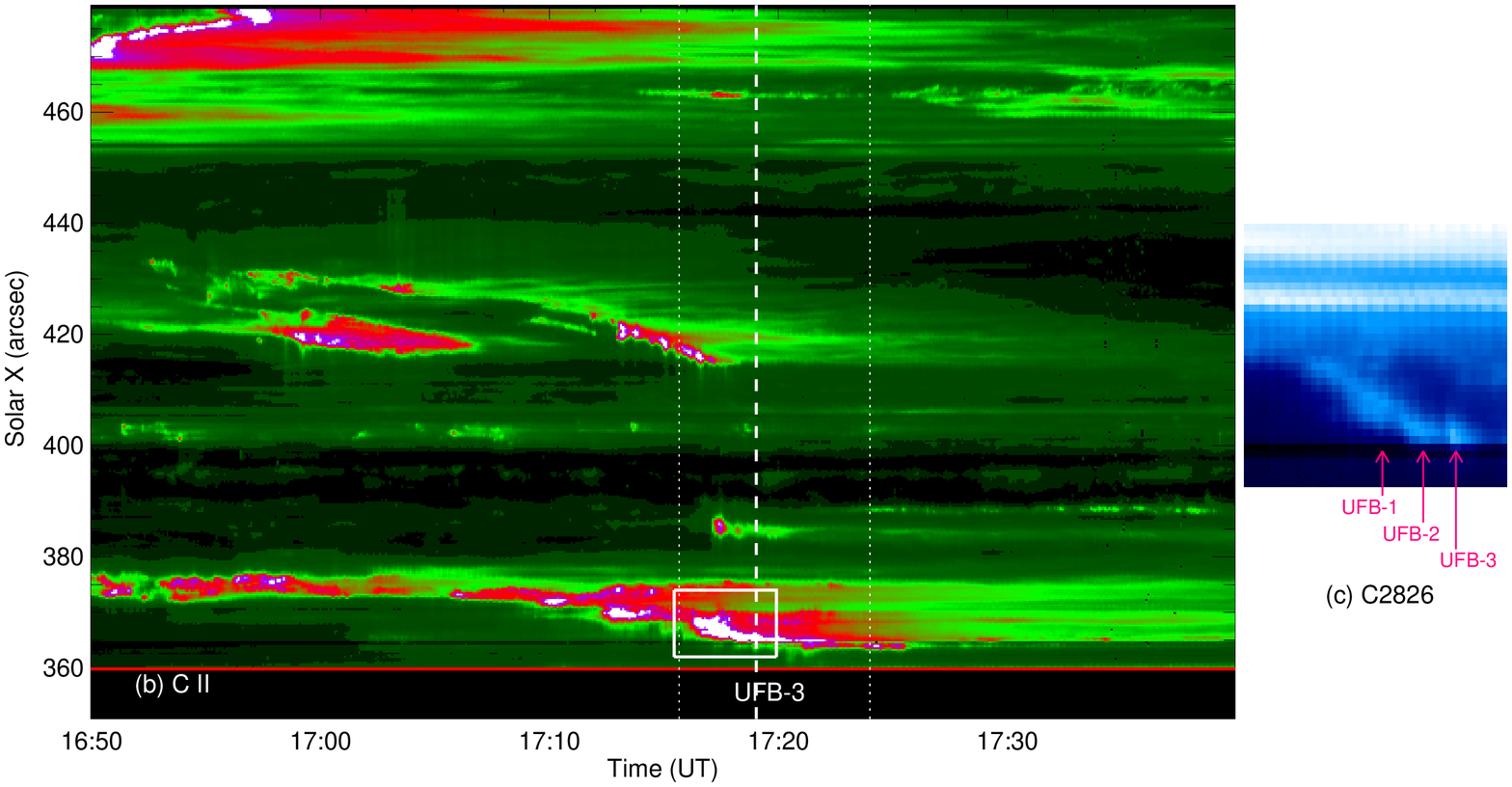}
\caption{    \textbf{(a)}  Fermi/GBM hard X-ray light curve (solid curve; from detector n4) and GOES soft X-ray light curve (dashed) of the X1 flare with GOES peak at 17:08.   The fourth major impulsive peak (indicated by vertical dotted lines) occurs at 17:17.  The vertical dashed line indicates the time of UFB-3 at 17:19 (see text).  \textbf{(b)} Evolution of the C II line-integrated intensity (linear scaling without continuum or pre-flare subtraction) from the IRIS FUV spectra.   The vertical lines are the same as in panel (a).   The red horizontal line indicates the positions below where there are no IRIS data.   \textbf{(c)} The NUV continuum intensity measure C2826 over a shorter time interval (17:15:27 to 17:19:55; x-axis) and smaller spatial extent along the slit (solar X from 362\arcsec\ to 374\arcsec), corresponding to the boxed region in panel (b).  The umbral flare brightenings are indicated as UFB-1, UFB-2, and UFB-3, and the intensity scaling is logarithmic.   The horizontal dark lines are the fiducial mark. }   \label{fig:context}
\end{figure}

The Fermi/GBM data from detector n4 at hard X-ray energies, $E=58-72$ keV, are shown in Figure \ref{fig:context}(a), and the light curve exhibits four main hard X-ray peaks at 16:59:17, 17:00:23, 17:03:31, and 17:17:27 with comparable amplitudes\footnote{The main peak at 17:04 may suffer from some pileup, thus making the 58-72 keV light curve an upper limit on the flux during this peak.}.   Also shown are the soft X-rays ($1-8$ \AA)  from GOES\footnote{\citet{Bamba2017} claim there is an error in the time-registry for these GOES data, but we do not see evidence to support this.}.

\begin{figure}[h!]
\centering
\includegraphics[scale=0.4]{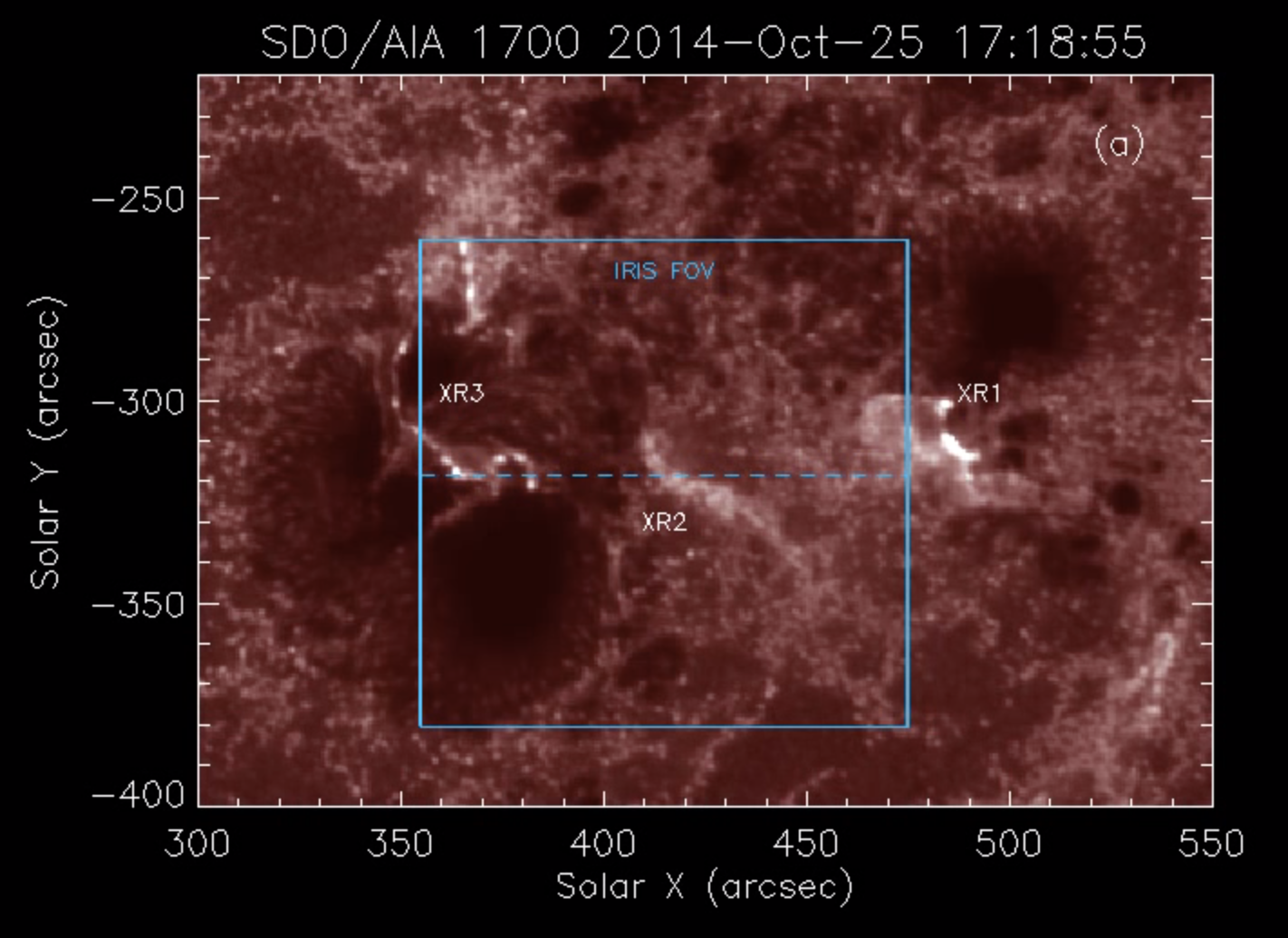}
\includegraphics[scale=0.4]{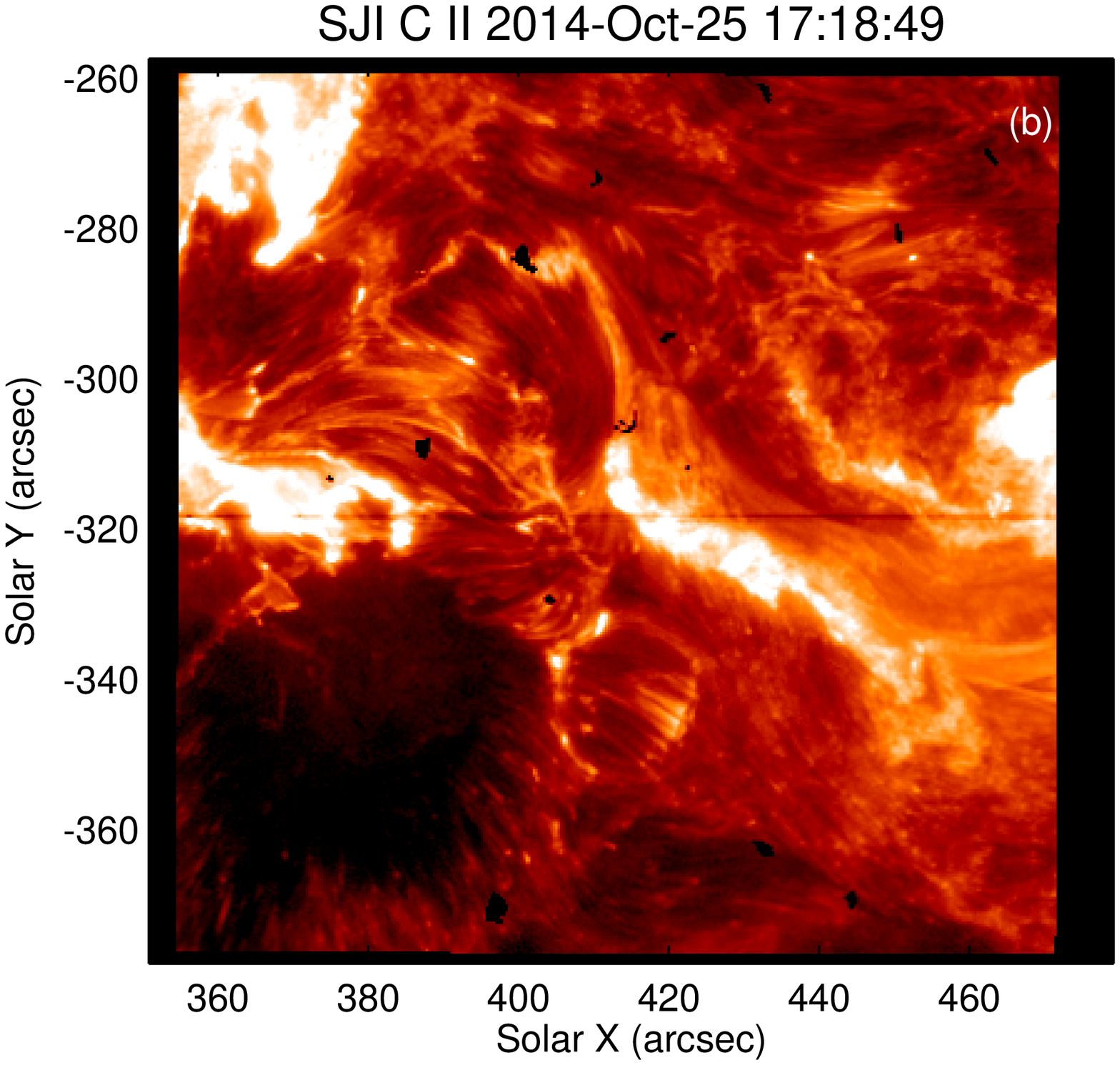}
\includegraphics[scale=0.4]{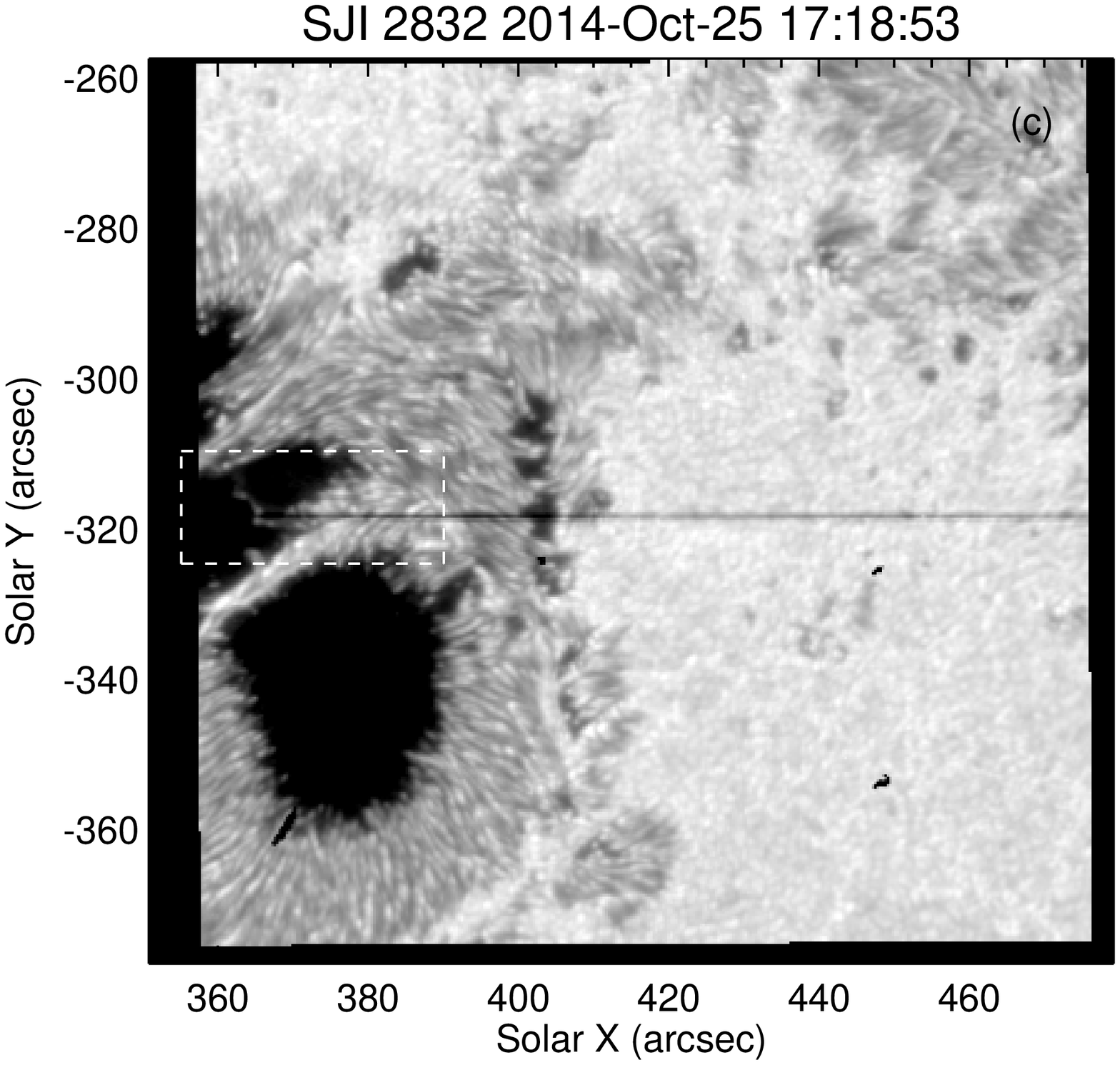}
\vspace{-15mm}
\caption{   IRIS slit context images of the flare ribbons at late times in the 2014-Oct-24 X1 flare in three different wavelengths.  \textbf{(a)} SDO/AIA 1700 image (linear intensity scaling) with the XR1, XR2, XR3 ribbons labeled following the nomenclature in \citet{Bamba2017}.  The IRIS slit is indicated as a horizontal dashed line.  \textbf{(b)} IRIS slit jaw image in the FUV (C II) with a logarithmic intensity scaling.   \textbf{(c)} IRIS slit jaw image in the Mg II h wing (SJI 2832) with a logarithmic intensity scaling.  The box indicates the field-of-view over which the area is calculated in Section \ref{sec:area}.  The XR3 ribbon crossing the IRIS slit in the umbra at solar $x \sim 365$\arcsec\ is apparent at this time (at the onset of the rise phase of UFB-3; see text) in all images.    }   \label{fig:context_sjis}
\end{figure}

The SOL2014-10-25T17:08:00 flare was a three-ribbon flare with two ribbons in the sunspot plage and a third ribbon 
crossing two umbrae within a triple-umbral complex \citep{Bamba2017}.  The X1 flare was the largest and the last of three flares that were triggered by an intruding positive polarity to the 
east of the main polarity inversion line, in accord with the triggering mechanism of \citet{Kusano2012}.  
Following the nomenclature for the ribbons used in \citet{Bamba2017}, the XR3 ribbon develops into the umbrae and crosses the IRIS slit during the decay phase of the fourth Fermi/GBM hard X-ray peak in Figure \ref{fig:context}(a).  
 The XR3 ribbon is seen in Figure \ref{fig:context_sjis}, where we show three context images at 17:19:  panel (a) shows the Solar Dynamics Observatory (SDO)/Atmospheric Imaging Assembly \citep[AIA; ][]{Lemen2012} 1700 image with the IRIS field-of-view (FOV) indicated, panel (b) shows an IRIS SJI C II image, and panel (c) is the IRIS SJI 2832 image.  The XR3 ribbon is very faint in SJI 2832, but the contrast against the umbral intensity is large.   \citet{Li2018} found Fe XXI emission blueshifted by 143 km s$^{-1}$ in the IRIS/FUV spectra of the XR3 ribbon.  The high temperature, low-density plasma are not analyzed further here.  
A comparison of flare intensity in the IRIS NUV spectra and  SJI 2832 images for the X1 flare has been discussed in \citet{Kleint2017}, who found that the relative contribution of Balmer continuum emission to the SJI 2832 radiation is nearly 70\% during the times when the XR3 ribbon crosses the slit (see the middle panels of their Figure 6).  In this paper, we present a detailed analysis of the IRIS spectra of the umbral flare ribbon XR3.

\section{A New Diagnostic for Deep Atmospheric Heating} \label{sec:irisanalysis}
\subsection{Identification of Near-UV Continuum Flare Kernels}

In this section, we present the C II ribbon development and describe how we identify the largest NUV continuum enhancements, which occur in the late phase when the XR3 ribbon develops into an umbra. The large contrast of the flare ribbons in umbrae (which are rather unusual environments for flare ribbon development) provide more robust comparisons to RHD models than flare spectra with low intensity contrast in the plage regions, which vary significantly due to the granulation evolution. 

The flare ribbons crossed the slit at many times and locations in the 2014-Oct-25 X1 flare.  The intensity integrated over each C II line from $-60$ to $+90$ km/s is shown in Figure \ref{fig:context}(b) and readily indicates the times when, and locations where, the XR1, XR2, and XR3 flare ribbons cross the IRIS slit.  However, the detailed formation of C II in flares has yet to be investigated for a range of heating levels outside of quiet-Sun conditions \citep{Rathore2015}.  Since C II has previously been analyzed extensively in IRIS data of solar flares \citep[e.g.,][]{Tian2015, Sadykov2016}, and they provide a comprehensive, high-contrast overview of the flaring ribbon evolution, we show their line-integrated evolution here but do not return to a detailed analysis of these lines in this paper.

 We identify candidate NUV continuum enhancements (and by proxy, white-light-emitting kernels) with a running difference of the line center intensity of Fe \textsc{ii} at $\lambda = 2814.45$ \AA, which we expect to brighten with the NUV continuum due to their similar formation conditions in flares \citep{Kowalski2017A}. Detailed justification for using Fe II to find candidate continuum enhancements is given in  Appendix \ref{sec:feii}.
 The time-evolution of the NUV continuum region from $\lambda=2824.5$ \AA\ to 2825.9 \AA\ is shown in Figure \ref{fig:context}(c) over a limited time range.  Following \citet{Kowalski2017A}, we calculate the average intensity in this wavelength window as C2826\footnote{The value of C2826 was defined as the average intensity from $\lambda=2825.6-2825.9$ \AA\ in the 2014-Mar-29 spectra in \citet{Kowalski2017A}.  In this study, we extend the blue end of the wavelength window for C2826 to 2824.5 \AA, which is line-free in the 2014-Oct-25 flare.}; the XR3 flare ribbon in C2826 crosses the IRIS slit and sunspot umbra from 17:17-17:19 UT in Figure \ref{fig:context}(c) and is further evident in SJI 2832 in Figure \ref{fig:context_sjis}(c).  Subtracting the preflare intensity gives the ``excess C2826'', hereafter C2826\prim\ (following the ``prime'' terminology for flare-only flux in \citet{Kowalski2012} and \citet{Kowalski2018}) and dividing by the pre-flare values gives the ``C2826 enhancement'' \citep[following the terminology in][]{Kowalski2015HSG}.  

The largest values of the C2826\prim\ (intensity) and the C2826 enhancement occur in the sunspot umbra at ($x, y$)$=$(364.8\arcsec, -318.4\arcsec) at 17:18:24 and 17:19:02 UT during the decay of the fourth hard X-ray event that peaks at 17:17:30.   
Hereafter, we refer to the brightest flaring pixels in the umbra during this hard X-ray event as the ``umbral flare brightenings'' (UFBs).  The time of UFB-3 is indicated in Figure \ref{fig:context}(a) for reference.  The peak of UFB-1 occurs at 17:17:31 at the spatial location of ($x,y = 365.8, -318.4$)\arcsec.  The peaks of UFB-2 at 17:18:24 and UFB-3 at 17:19:02 UT occur in nearly the same location at ($x,y = 364.8, -318.4$)\arcsec.   However, UFB-2 appears to be less than one pixel to the west (in the positive $x$ direction) of UFB-3, and the apparent motion of the ribbon eastward along the slit is evident from the time of UFB-1 to the time of UFB-3 (see Figure \ref{fig:context}(c)).  
Hereafter, we focus our analysis on the spectra of UFB-2 and UFB-3, which produced the largest and fastest NUV continuum enhancements in the IRIS spectra.

 \subsection{Continuum Properties of the Umbral Flare Brightenings}
\begin{figure}[h!]
\centering
\includegraphics[scale=0.75]{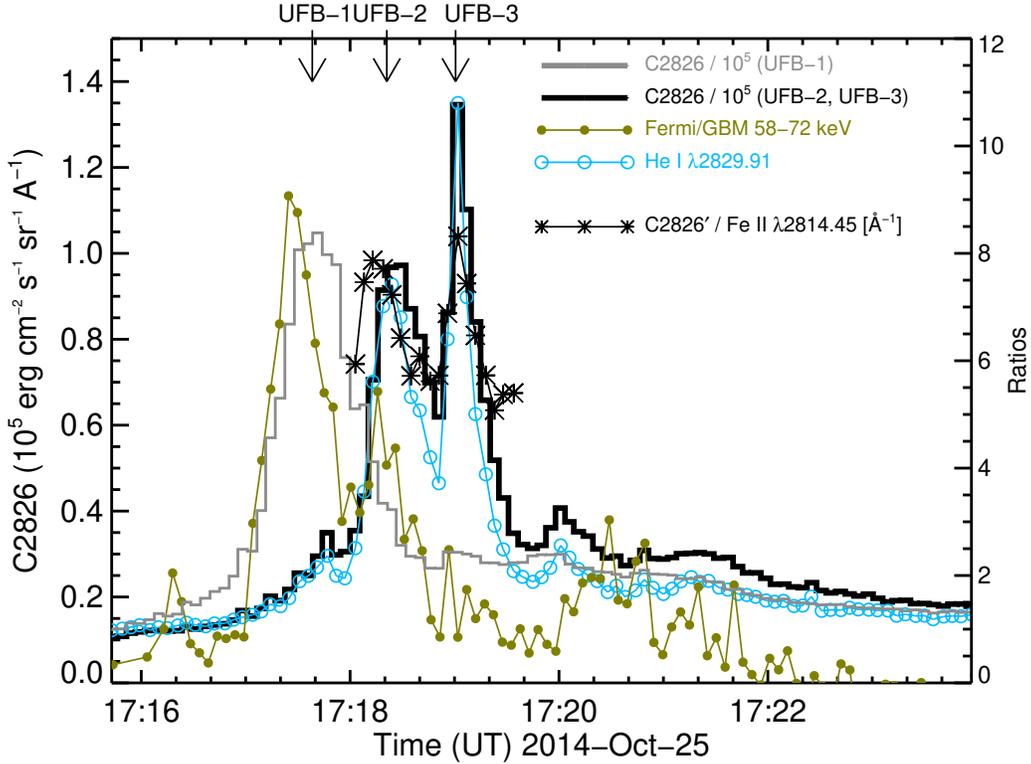}
\caption{ C2826 light curve of the umbral flare brightenings UFB-2 and UFB-3 at ($x,y$)$ = $(364.8, -318.4) arcsec during the fourth major hard X-ray peak, over the same time range as indicated by the vertical dotted lines in Figure \ref{fig:context}(a,b).  The peaks of UFB-2 and UFB-3 are delayed with respect to the peak of the hard X-rays in Fermi/GBM (olive) because the Fermi/GBM is not spatially resolved at the Sun and the ribbons cross the IRIS slit at this spatial location at a later time.  The continuum-to-line ratios (C2826\prim /\feii; asterisks, right axis) exhibit very large values of $\sim7-8$ in the impulsive phases (from 17:18:03 to 17:19:34) of UFB-2 and UFB-3.    The light curve of C2826 at the spatial location of UFB-1 is shown in a  lighter gray tone; UFB-1 reaches its maximum closer in time to the HXR peak.   }   \label{fig:lc}
\end{figure}

The time-evolution of C2826 averaged over three spatial pixels (the spatially averaged intensity, $<I_{\lambda}>$) centered at ($x, y$)$=$(364.8\arcsec, -318.4\arcsec) is shown in Figure \ref{fig:lc}.  The UFB-2 and UFB-3 events correspond to the first and second C2826 peaks, respectively.  The C2826 enhancement remarkably exceeds the  pre-flare umbral intensity by a factor of ten.
The $E=58 - 72$ keV X-ray light curve is overplotted in Figure \ref{fig:lc}, and UFB-3 occurs when the ribbons cross the slit in the decay phase of this hard X-ray peak. From the range of calculated FWHM values of the C2826\prim\ light curve of UFB-3, we constrain upper limits on the duration of heating to $\Delta t_{\rm{burst}} = 12 - 22$~s over a solid angle of 0.33\arcsec\ (slit width) $\times$ (0.40\arcsec; NUV spatial resolution), or an area of $\sim6.5 \times 10^{14}$ cm$^{2}$.  However, the data do not exclude shorter heating durations over smaller areas.  Furthermore, the rise phase of UFB-3 is unresolved since it consists of one temporal point.  However, we note $10-20$~s is generally the timescale range for soft X-ray derivative bursts characterized in another flare \citep{Rubio2016}.  Several heating episodes may occur (at least partially) in the same location (e.g., UFB-2 and UFB-3), but this cannot be quantified in any detail even with the impressive spatial resolution of IRIS since the ribbons move along the slit from the time of UFB-1 to the time of UFB-3.  

\begin{figure}[h!]
\centering
\includegraphics[scale=0.5]{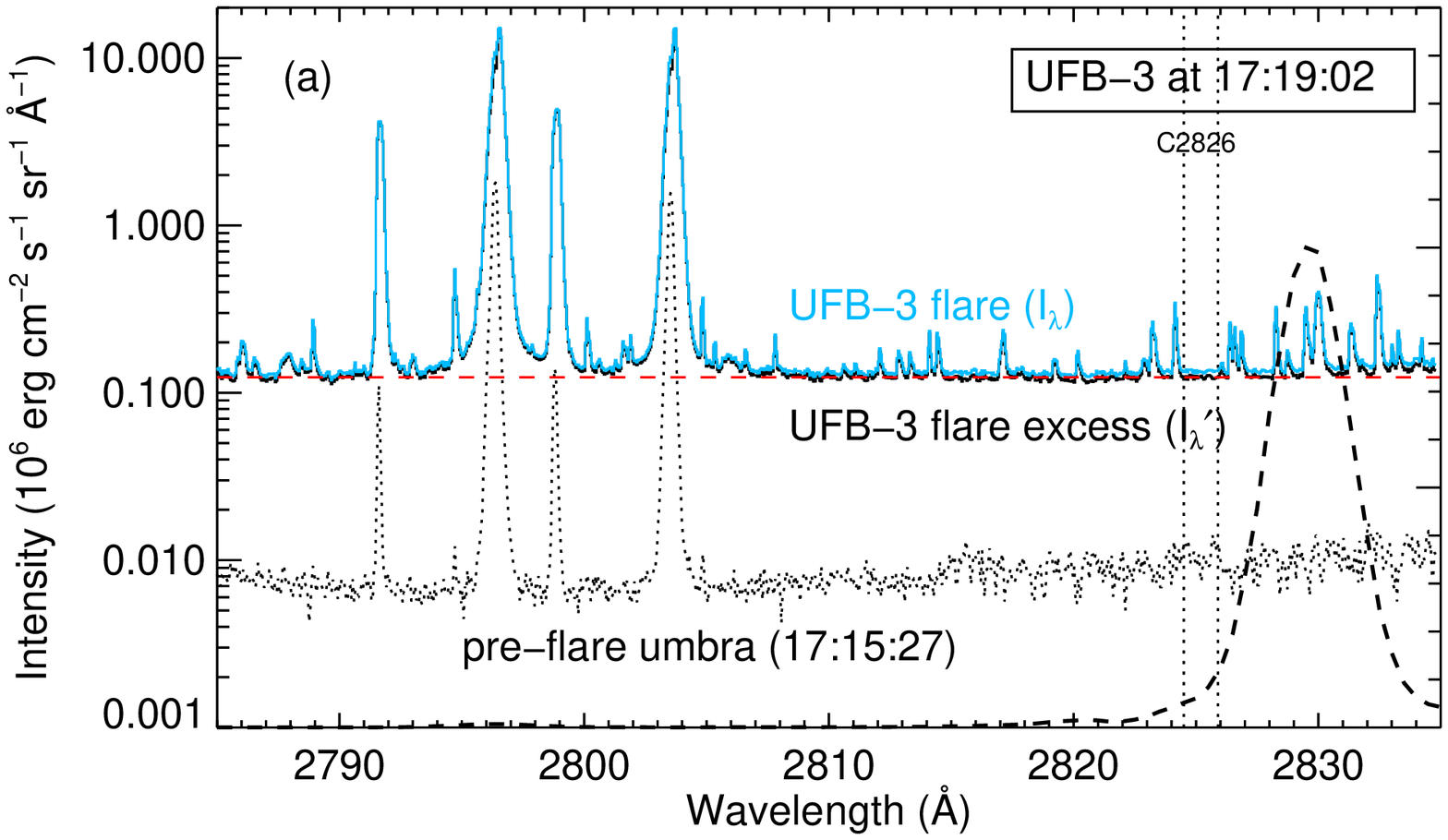}
\includegraphics[scale=0.5]{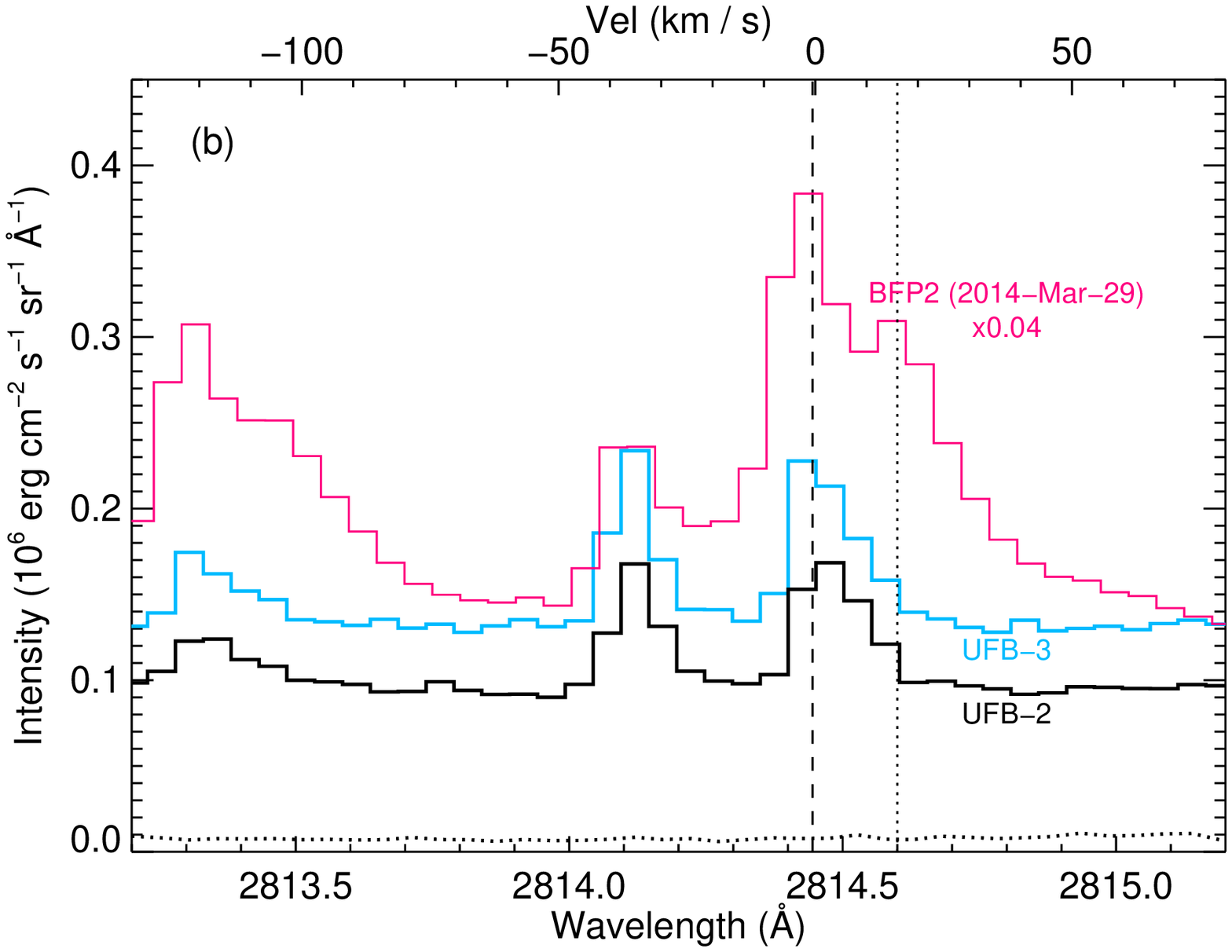}
\caption{ \textbf{(a)} IRIS NUV spectra of the brightest peak of the umbral flare brightening light curve in Figure \ref{fig:lc}.  The spectra were averaged over three pixels (1\arcsec) in the spatial dimension.  Vertical dotted lines show the extent of the line-free continuum region C2826.  The horizontal red dashed line indicates the value of C2826\prim\ extrapolated over the full spectral range of the IRIS NUV; C2826\prim\ adequately represents the NUV continuum flare excess intensity over this range.  The effective area curve for the IRIS slit jaw SJI 2832 is shown as a dashed curve for reference.  \textbf{(b)} The \feii\ lines in the UFB-2 (black) and UFB-3 (light blue) spectra indicate asymmetries at redder wavelengths than the rest wavelength, which is indicated by the vertical dashed line.  The redmost extent of the \feii\ line in UFB-2 and UFB-3 is indicated by a vertical dotted line.  The other flare emission lines in panel (b) are Fe II at $\lambda=2813.322$ \AA\ and Fe I at $\lambda=2814.115$ \AA. The umbral flare brightening profiles are compared to the spectra (scaled by 0.04; see Section \ref{sec:discussion}) of the bright flare footpoint \citep[``BFP2'';][]{Kowalski2017A} in the X1 flare on 2014-Mar-29.   The 2014-Mar-29 flare data have been binned to the same spectral binning as the 2014-Oct-25 flare data and are shown averaged over three pixels in the spatial direction.  The red-wing asymmetry  in the 2014-Mar-29 flare is broader, brighter, and redder; it is also clearly spectrally resolved when this binning is not applied. }   \label{fig:spectrum}
\end{figure}

The IRIS NUV spectrum at the peak of UFB-3 is shown in Figure \ref{fig:spectrum}(a).  This spectrum exhibits the most robust characterization of the IRIS NUV continuum to date during a solar flare.  The flare spectrum shows that the NUV continuum intensity level extrapolated (over the horizontal red line) from C2826\prim\ extends over a spectral range of $\Delta \lambda = 50$ \AA.  The horizontal line extrapolation provides an estimate for the relative, wavelength-integrated continuum energy in the NUV spectral range of IRIS.  We find that only $\sim25$\% of the UFB-3 peak flare kernel brightness in the IRIS NUV spectral range is attributed to the continuum radiation, while the majority of the wavelength-integrated kernel brightness (nearly 70\%) is due to the Mg II lines.

The C2826\prim\ intensity from IRIS during other flares has been used to compare directly to radiative-hydrodynamic flare models \citep{Heinzel2014, Kleint2016, Kowalski2017A}. 
But are the C2826\prim\ ribbons in the 2014 Oct 25 X1 flare spatially resolved for a direct comparison of the value of C2826\prim\ ($<I_{\lambda}'> \approx 1-1.5\times10^5$ erg s$^{-1}$ cm$^{-2}$ sr$^{-1}$ \AA$^{-1}$; Figure \ref{fig:lc}) to model snapshots?  We use the \emph{iris\_sg\_deconvolve.pro} routine to deconvolve the UFB-1 and UFB-3 peak spectra by the IRIS PSF as described in \citet{deconv}.  We find that after $15$ iterations of the deconvolution procedure, 
the NUV spectra increase in brightness by factors of $1.2-2$. 
The IRIS NUV resolution is 0.4\arcsec\ or 1.2 (binned) pixels, and the spatial full-widths-at-half-maxima of UFB-1, UFB-2, and UFB-3\footnote{UFB-3 occurs adjacent to a fiducial mark in the spectra (Figure \ref{fig:context}b), and we are unable to quantify the full spatial extent of the kernel.} are 3-6 pixels, corresponding to $700-1400$ km at the Sun;  in Figure \ref{fig:context}(c), it is apparent that the NUV continuum radiation  (C2826) from the UFBs is bright over a few pixels only.   In higher resolution H$\alpha$ data of umbral flare ribbons, very narrow widths of $\sim100$ km are observed \citep{Sharykin2014}, which  suggests there may be significant unresolved structure in our IRIS observations.  
Higher-quality observations (i.e., without saturation) of the widths and motions of plage and umbral kernels will be important to obtain with the Daniel K. Inouye Solar Telescope at very high time resolution.

 \subsection{The Fe II 2814.45 \AA\ line in umbral flare brightenings} \label{sec:lines}
 
The \feii\ emission line has been recently studied with radiative-hydrodynamic models of flares \citep{Kowalski2017A} and has revealed two emission line components:  a line component centered near the rest wavelength and a spectrally resolved, broad emission component to the red of the rest wavelength.  Compared to other chromospheric flare lines, the \feii\ line exhibits one of the lowest opacities, and the optical depth does not build up as fast in that line as in the Balmer lines or in Mg II.  Thus, the properties of \feii\ constrain the dynamics in deep layers of the flaring chromospheric condensation and stationary flare layers below \citep{Kowalski2017A}, where the white-light and IRIS NUV continuum radiation may originate.   The similar temperature and density dependencies of \feii\ and hydrogen Balmer bound-free radiation are shown in Appendix A.  

In Figure \ref{fig:spectrum}(b), we show an inset of several flare emission lines around the region of the \feii\ line in UFB-2 and UFB-3.   We measure the bisector at 30\% maximum for the \feii\ \AA\ emission line to be $+3.7\pm2.9$ km s$^{-1}$.  This is just greater than a 1$\sigma$ redshift detection\footnote{Adding a systematic uncertainty of 1.8 km s$^{-1}$ and a statistical uncertainty of 2.3 km s$^{-1}$ in quadrature.}.
However, the line profile shape exhibits asymmetric brightening at $\lambda > \lambda_{\rm{rest}}$ such that there are more spectral bins with an intensity greater than or equal to the line half-maximum on the red side than on the blue side.  We show vertical dashed lines at $\lambda_{\rm{rest}}$ and vertical dotted lines at $\lambda-\lambda_{\rm{rest}} = +16$ km s$^{-1}$ to indicate the red extent of this profile asymmetry.  The Fe II at $\lambda = 2813.32$ \AA\ line exhibits a similar asymmetric profile and redshift ($4.5$ km s$^{-1}$) to the Fe II at $\lambda=2814.45$ \AA\ line, while the Fe I $\lambda=2814.115$ \AA\ line exhibits no evidence of a redshift or an asymmetric profile.  Furthermore, UFB-2 and other flare spectra show similarly redshifted and asymmetric \feii\ profiles to UFB-3.  
 These three lines are of particular utility for studying flares, as they fall within a standard window common to the IRIS NUV line lists, and thus provide the opportunity to compare with a large archive of observations.  With full spectral readout, there are numerous additional Fe II and Cr II lines with similar profiles available, as discussed in Section~\ref{sec:lines} and Appendix B. The asymmetric, redshifted profiles of these Fe II lines and other emission lines can be synthesized from RHD models to constrain the chromospheric velocity field in white-light emitting layers at high-time resolution.

\subsection{The continuum-to-line ratio in IRIS spectra} \label{sec:c2l}
The umbral flare brightening spectra reveal that the ratio of the NUV continuum radiation to  the Fe II line intensity is an interesting diagnostic in flares.  This continuum-to-line ratio can be readily obtained from spectra with narrow wavelength coverage; like an equivalent width, it is independent of uncertainties \citep[e.g.,][]{Wusler2018, deconv} in the absolute intensity calibration. 
We define the continuum-to-line ratio as the excess C2826 (C2826\prim) divided by the wavelength-integrated, continuum-subtracted, and preflare-subtracted intensity of the \feii\ line.  The continuum-to-line ratio is hereafter denoted as C2826\prim/\feii\ (with units of \AA$^{-1}$). The evolution of this quantity varies between $7 - 8$ over the peaks of UFB-2 and UFB-3 and is shown in Figure \ref{fig:lc}. At the peak of UFB-1 (not shown), the continuum-to-line ratio is also large near $7$.   Subtracting the last decay phase spectrum of UFB-2 from the peak of UFB-3 at the spatial location of the light curve in Figure \ref{fig:lc} gives the newly-brightened flare radiation assuming that UFB-2 and UFB-3 occur at different locations within the spatial resolution.  For this spectrum, the value of C2826\prim/\feii\ is even larger ($\approx12$).  A comparison of the C2826\prim/\feii\ ratios among many flares will be presented in Butler et al.\ (2019, in prep.).

The \feii\ line in the IRIS NUV spectrum is an important diagnostic in flares because the line is produced (assuming LTE) over a similar temperature range of $T\sim8000-18,000$ K as hydrogen bound-free radiation \citep[][and shown in detail in Appendix A]{Kowalski2017A}, which contributes to the C2826\prim\ intensity in flares \citep{Heinzel2014}. Thus, the ratio of these two quantities
 reflects the amount of significant heating (T$\gtrsim8500$ K) in  deep layers of the atmosphere that exhibit $\tau_{\lambda} \lesssim 1$ for continuum radiation and $\tau_{\lambda} \gtrsim 1$ for \feii\ radiation. 
The ratios from optically thin, uniform slabs with low to moderately high densities ($\rho < 10^{-9}$ g cm$^{-3}$) at $T\sim10,000$ K are $\lesssim0.8$ (Appendix A), which are far less than the observed values of $7-8$ in the UFBs.   Radiative-hydrodynamic modeling and the detailed interpretation and analysis of large values of C2826\prim/\feii\ in the UFBs are outside the scope of this paper and will be presented in Paper II.

\section{Other Observational Properties of the Flare} \label{sec:otherstuff}
In order to model the UFBs and the continnum-to-line ratios with electron beam heating simulations in Paper II, several comprehensive input parameters (electron beam energy and power-law index) and other constraints from the emission lines are presented in this section.  In Section \ref{sec:area}, we calculate the flare area from high spatial resolution imagery of the flare ribbons.  This flare area is combined with the nonthermal electron power inferred from Fermi/GBM data in Section \ref{sec:thicktarg} to give the energy flux density of beams assumed to heat the lower atmosphere.  In Section \ref{sec:lines} and appendix~\ref{sec:lineidsapp}, we discuss properties of the many other Fe II emission lines in the UFBs, as well as species such as Cr II and Fe I, and the constraints they provide on atmospheric models.  In Section \ref{sec:heliumI}, we discuss the properties of a He I emission line in the IRIS NUV that will complement the constraints from the Fe II lines.

\subsection{Evolution of the Flare Ribbons}\label{sec:area}
High spatial resolution flare area estimates are critical for obtaining an accurate heating rate for modeling the RHD response \citep{Krucker2011}.  However, the continuum contrast outside the umbra is low, which causes flare area measurements to be difficult in solar flares for even large flare energies.  
Generally, the non-flaring photospheric intensity in the NUV Mg II h wing is 50x brighter than the pre-flare umbral NUV intensity, and the large amplitude, gradual variation from photospheric convection outside of the sunspot makes the identification of bona-fide flare continuum enhancements difficult \citep[we refer the reader to][for some properties of NUV enhancements away from the sunspots]{Kleint2017}.   

We present an algorithm that identifies pixels that brighten impulsively in flares.  This algorithm excludes gradual evolution of the solar granulation that can be falsely identified as flare kernels, and the method is applicable to other data sets with high-time and high-spatial resolution.   Using the intensity-calibrated SJI 2832 images (with a cadence of 16~s), we set a threshold excess intensity (T.E.I.) value for impulsive pixels in SJI 2832 to $10^5$ \intunits, which corresponds to $\sim30$\% of the maximum excess intensity observed over the entire observation sequence.  We consider here only the umbral region indicated in the SJI 2832 image by the box in Figure \ref{fig:context_sjis}(c).  A pixel at time $t$ in image $i$ is counted as an impulsive flare pixel if the following criteria are met:
1) the excess intensity in this pixel at $t$ is greater than T.E.I.; 
2) the excess intensity in this pixel at $t - 48$~s ($i-3$) or $t-64$~s ($i-4$)  is less than 0.5 x T.E.I; and 3) the excess intensity at $t+160$~s ($i+10$) is less than 0.5 x T.E.I.  These criteria identify newly flaring, impulsive pixels in SJI 2832 by efficiently excluding a significant number of pixels that exhibit a much more gradual evolution with sustained excess values.  It is not possible to tell whether these gradual brightenings are due to changes in the granulation or are related to the flare, as in the so-called ``type II white-light flare'' variations \citep{Matthews2003, Ondrej2017}.  Hereafter, we refer to the impulsive flare areas using this algorithm as ``newly-brightened'' flare areas.

\begin{figure}[h!]
\centering
\vspace{-75mm}
\includegraphics[scale=0.6]{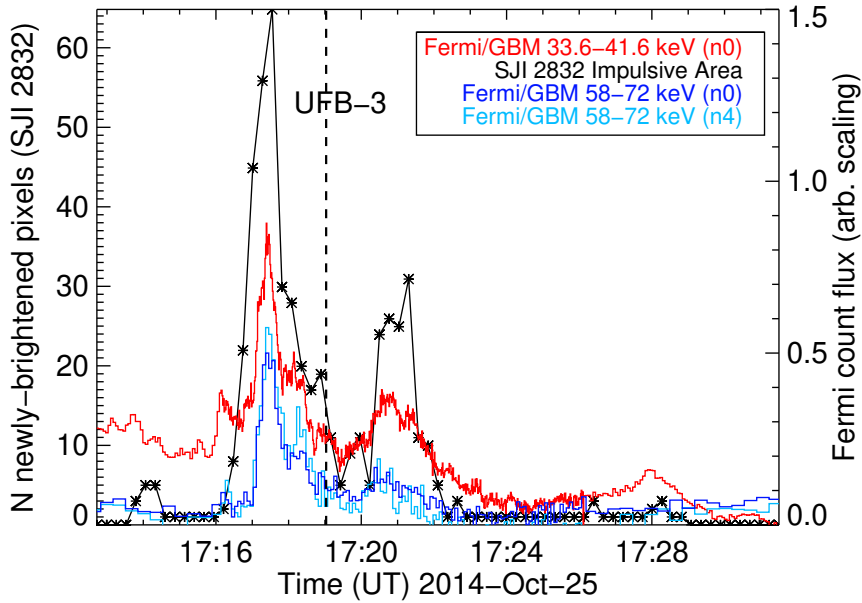}
\caption{   The newly-brightened (impulsive) flare area in SJI 2832 and the Fermi/GBM hard X-ray light curve during the fourth hard X-ray peak.  The conversion from pixels to area for these observations (which employ on-board binning) is $6\times10^{14}$ cm$^2$ / pixel.  The UFB-3 occurs during the late phase of the fourth hard X-ray peak, but the evolution of the flare area in SJI 2832 follows the hard X-ray light curves.   Note, the newly-brightened flare area in SJI 2832 is calculated from the region indicated by the box indicated in panel (c) of Figure \ref{fig:context_sjis}. }   \label{fig:sji_lc}
\end{figure}

The number of newly-brightened flare pixels (where each pixel corresponds to an area of $6\times10^{14}$ cm$^{2}$ at the Sun) in the umbral region of SJI 2832 is shown as a function of time in Figure \ref{fig:sji_lc} and is compared to the Fermi/GBM $E=35-41$ keV and $E=58-72$ keV light curves.  The newly-brightened flare area and hard X-ray emissions follow each other closely (see Graham et al., in prep), which suggests that the heating in these pixels (over a small region in the flare) is related to the non-thermal electrons that produce the fourth hard X-ray peak\footnote{This also shows that the fourth hard X-ray peak in Fermi/GBM is a result of the X1 flare ribbons and  not due to a particle event with a non-solar origin.}.    In the decay phase of the fourth hard X-ray peak, the newly-brightened ribbon areas have decreased as they progress through the umbra.  Thus, there are \emph{impulsive} flaring pixels in the \emph{gradual} decay of the hard X-ray peak (which occurs in the gradual decay of a much larger soft X-ray peak). 
From Figure \ref{fig:sji_lc}, we obtain a range of newly-brightened flare areas to be $7\times10^{15}$ cm$^2$ (17:19; at the peak UFB-3) to $4\times10^{16}$ cm$^{2}$ (17:17:30; at the peak of the fourth HXR peak).    This information is combined with constraints from the Fermi/GBM spectrum in Section \ref{sec:thicktarg} to infer the energy flux in electron beams.

Whereas the Fermi/GBM hard X-rays originate from the entire flaring area, the SJI 2832 data used to calculate the flare area are rather limited in field of view (see Figure \ref{fig:context_sjis}(c)).  
Thus, we use our algorithm\footnote{To calculate newly-brightened emission in SDO/AIA 1700 at time $t$ we used running differences of $t-24$~s, $t-48$~s, and $t+168$~s, and we added the number of pixels that were below 50\% of a high threshold value, which we took to be 30\% of the saturation ($\sim15,000$ counts).} to calculate the newly-brightened flare area in SDO/1700, which provides a complete field-of-view characterization of the flare region but with lower spatial resolution.  The newly-brightened flare area from SDO/1700 is  $\sim2\times 10^{17}$ cm$^{2}$ at the time of the peak of the fourth hard X-ray light curve, and it falls to $\sim5\times 10^{16}$ cm$^{2}$ at the times of the (spectral) UFBs.  These areas are 10-40x larger than obtained from the umbral region in SJI 2832.  Note that several bright kernels in the SDO images are saturated during the flare, making the areas from SDO/1700 over-estimates of the true flaring area at some times.

We also investigate whether the bright flare sources that cross the IRIS slit  (UFB-1, UFB-2, or UFB-3) are obviously brighter than other newly-brightened flare pixels in the SJI 2832 images within the box indicated in Figure \ref{fig:context_sjis}(c).  
Many of the brightest SJI 2832 pixels occur near the maximum of the HXR light curve in Figure \ref{fig:sji_lc}.   To determine how the UFBs from the spectra compare to the brightest flare excess pixels that do not cross the slit, we fold the IRIS spectra at the peaks of UFB-2 and UFB-3 with the effective area curve of SJI 2832 (Figure \ref{fig:spectrum}) to produce the synthetic values of SJI 2832; the synthetic excess intensity values\footnote{The synthetic SJI 2832 intensity values from the UFB spectra are not averaged over three spatial pixels in this analysis.} are $1.7\times10^5$ and $2.3\times10^5$ erg cm$^{-2}$ s$^{-1}$ sr$^{-1}$ \AA$^{-1}$ for UFB-2 and UFB-3, respectively, while the brightest SJI 2832 excess value is $3.5\times10^5$ erg cm$^{-2}$ s$^{-1}$ sr$^{-1}$ \AA$^{-1}$.
There are 117 pixels (or 23\% of the pixels brighter than the T.E.I.) in the SJI 2832 images that become as bright as or brighter than the UFB-2 spectrum, and there are 35 (or 7\% of the pixels brighter than the T.E.I.) greater than or equal to the brightness of UFB-3. Clearly, a significant number of pixels that do not intersect the IRIS slit are brighter than the synthesized SJI 2832 values from the UFB-2 and UFB-3 spectra, suggesting that the continuum brightness of the umbral flare brightenings (UFB-2 and UFB-3) that cross the IRIS slit are not uniquely large.

\subsection{Fermi/GBM Spectral Analysis}  \label{sec:thicktarg}

\begin{figure}[h!]
\centering
\includegraphics[scale=0.65]{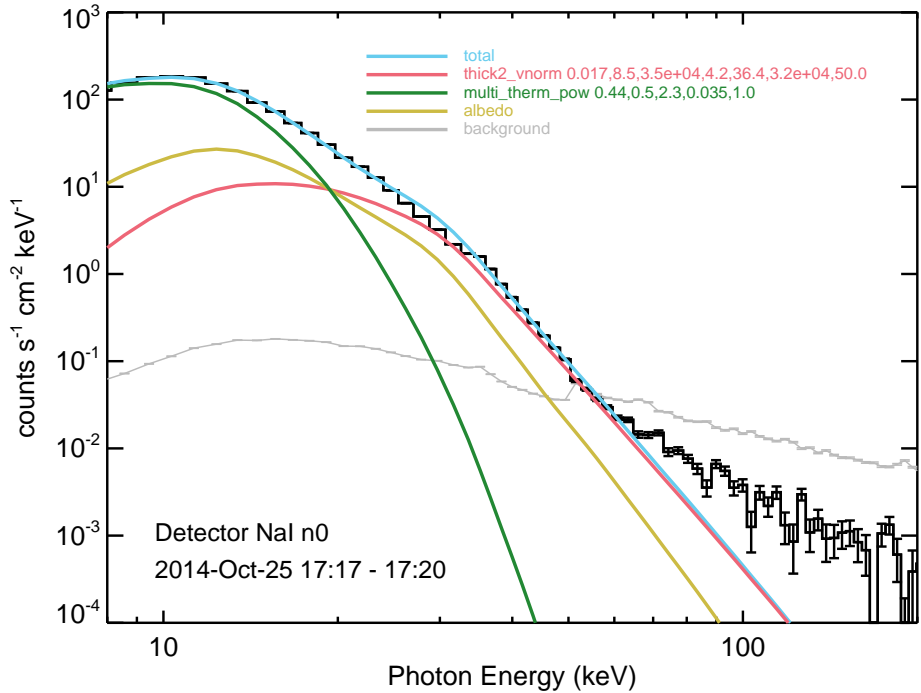}
\caption{  A multi-component fit to the $E=10-60$ keV Fermi/GBM (NaI n0 detector) spectrum from 17:17-17:20 with therm\_multi\_pow $+$ albedo $+$ thick2\_vnorm (``total'').  Parameters as returned by OSPEX for each component are given in the legend.  Fitting during times and using other methods of background subtraction give similar parameters.  Note, the $\sim20-40$ keV energy range is excluded from the fit due to known calibration issues with Fermi/GBM.    }   \label{fig:fits1}
\end{figure}

The hard X-rays from Fermi/GBM indicate the presence of nonthermal electrons in the flare, which are thought to produce the heating responsible for the white-light (e.g., C2826) increase.  To determine if nonthermal electrons (as constrained from Fermi) are sufficient to explain the IRIS spectra, we infer the properties of a nonthermal electron beam using the standard thick-target formulae. 
The nonthermal electron distribution is parameterized by a power-law index ($\delta$), an energy flux density (given by $a \times 10^x$  \fluxunits;  hereafter, abbreviated by $a$F$x$), and a low-energy cutoff ($E_c$) of the distribution, which is almost always an upper limit.  Following the thick-target modeling of \citet{Milligan2014}, we fit the Fermi/GBM spectrum at $E=10- 60$ keV with a sum of thick target bremsstrahlung emission, an albedo correction from Compton scattering off the photosphere \citep{Kontar2006}, and a multi-thermal power-law spectrum.  The fits  were done using the SolarSoft OSPEX software in IDL and are shown in Figure \ref{fig:fits1} for the time-interval 17:17 - 17:20 (indicated by vertical dotted lines in Figure \ref{fig:lc}).  Varying the fit interval to two minutes within this window gives similar results.   We fit all energy intervals at E $=10- 60$ keV using the same background times.  Because it is known that there is an iodine $k$-edge (at 33 keV)  in the detector \citep{Meegan2009} that causes spectral fitting from $20-40$ keV to be problematic, we also fit only $E=40-60$ keV with a thick target model but find a similar power law.  We conclude that the hard X-rays are consistent with a thick-target electron spectrum that has a power-law index of $\delta=8-9$ and an upper limit to the low-energy cutoff of $35-40$ keV.  Note, the Compton scattering and photo-electric absorption of outgoing hard X-ray photons from the heated chromosphere has not been taken into account in the X-ray spectral fits, whereas a limb-observation has shown that the  HXR source (in another flare) is low enough to make the Compton opacity significant \citep{Martinez2012}.  The extinction of hard X-rays in the evolved atmosphere will be considered in the radiative-hydrodynamic flare model atmospheres in Paper II.  

At $E\gtrsim$60 keV there is clearly a break in the spectrum in Figure \ref{fig:fits1}.  While there is a transient burst of hard X-ray and gamma ray photons at $E \gtrsim 100$ keV to $E \sim 10$ MeV that corresponds to the evolution of the C2826 light curve in Figure \ref{fig:lc}, this gamma-ray event cannot be constrained to originate from the Sun since Fermi was approaching the South Atlantic Anomaly at these times.  Thus, the break is likely due to variable background radiation.

 The last input parameter for RHD modeling is an estimate of the energy flux density ($a$F$x$) of accelerated electrons.   Using the fits in Figure \ref{fig:fits1}, we obtain a nonthermal power of $8\times10^{27}$ erg s$^{-1}$, which is an average over the time interval. A fit to the times of the fourth HXR peak at 17:17:20-17:17:40 gives a power of $10^{28}$ erg s$^{-1}$ (with similar cutoff and power-law indices).   The hard X-ray count rate at the time of UFB-3 is $\sim$1/10 of the max count rate in the fourth peak (see Figure \ref{fig:context}(a) and Figure \ref{fig:sji_lc}), which allows us to estimate that the nonthermal electron power is $10^{27}$ erg s$^{-1}$ at the time of the UFBs in the IRIS spectra.  The area estimated in Section \ref{sec:area} from SDO/1700 gives a flux density estimate of $2\times10^{10}$ \fluxunits, while the area from SJI 2832 gives $1.5\times10^{11}$ \fluxunits.
We thus obtain a range of 2F10$ - $2F11 for the nonthermal electron energy flux density for input to RADYN modeling.  

The power-law index of the hard X-ray emission is very soft at these late times in the flare.  The analysis of RHESSI data for this flare also indicates a large power-law index  at earlier times \citep{Kleint2017}.   The RHESSI imagery suggest the hard X-ray (25-50 keV) source is coronal, which may be due to a thick-target nonthermal coronal source or due to pileup of low-energy photons from a thermal coronal source.  Because Fermi/GBM is spatially unresolved, we cannot rule out that the $E\gtrsim40$ keV emission originates from a thick target coronal source \citep{Veronig2004}, but we have ensured that we use data from the detectors of Fermi that do not suffer from pileup of lower energy photons.  We note that \citet{Thalmann2015} used RHESSI data of other flares from AR 12192 and find very soft electron spectra of $\delta \sim 9$ in the decay phase and very large nonthermal electron energies as well. There are many other examples of soft power-law indices in flares in Solar Cycle 24 \citep{Milligan2014, Thalmann2015, Kerr2015, Warmuth2016A}.  

 \subsection{Other Emission Line Properties of the Umbral Flare Brightenings} \label{sec:lines}
\begin{figure}[h!]
\centering
\includegraphics[scale=0.75,angle=90]{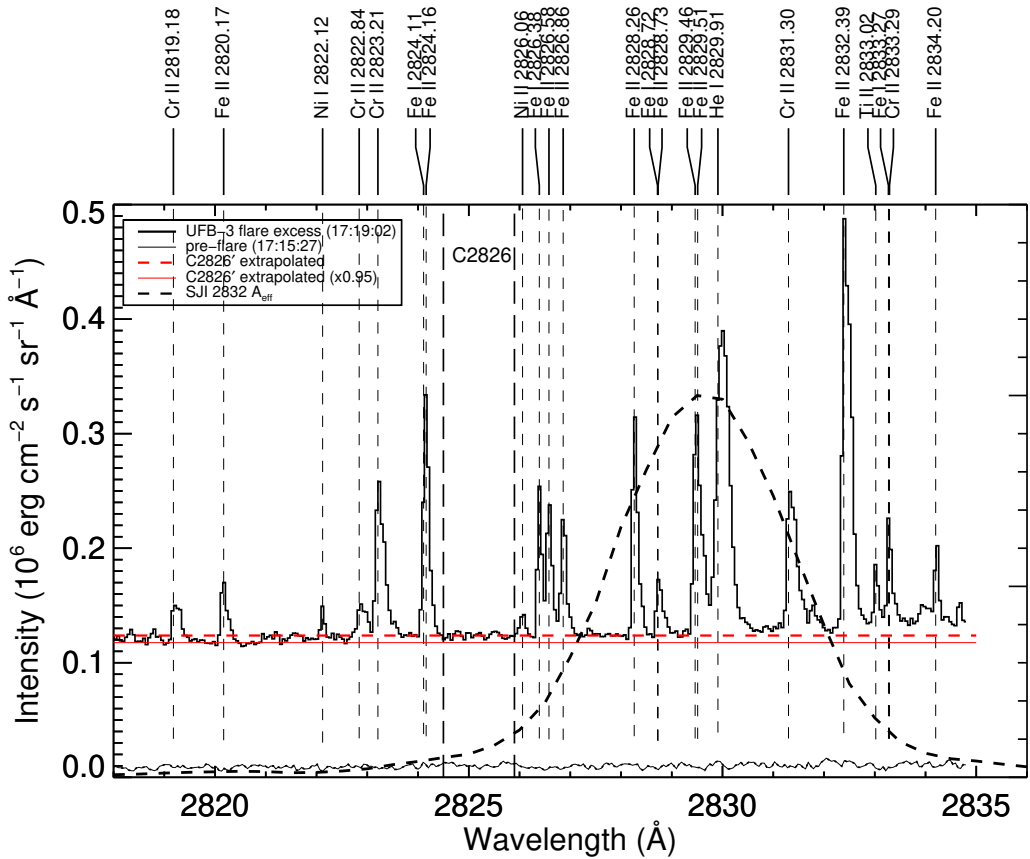}
\vspace{-5mm}
\caption{  Enlarged view of the UFB-3 peak flare spectrum over the wavelength range of SJI 2832.  Many Fe II and Cr II lines
exhibit similar profile asymmetries, and the He I 2829.91 \AA\ line is redshifted and broad.  We adjust the value of C2826\prim\ by 5\% to show that it better represents the excess NUV continuum intensity at $\lambda \lesssim 2823$ \AA.
}   \label{fig:sji_spec}
\end{figure}

\begin{figure}[h!]
\centering
\includegraphics[scale=0.75,angle=90]{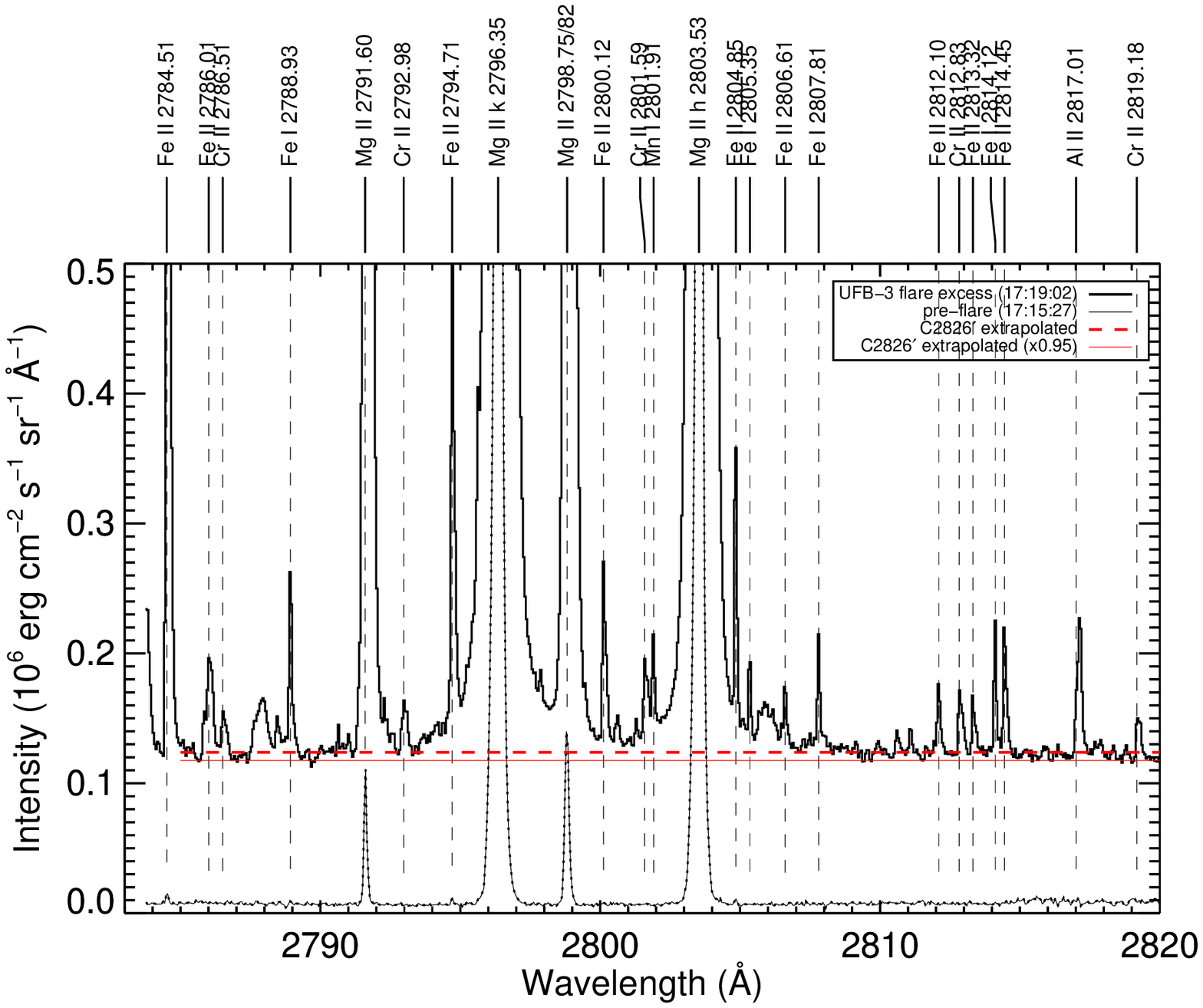}
\vspace{-5mm}
\caption{  Some line identifications of the brighter lines  in the UFB-3 peak flare spectrum at shorter NUV wavelengths around Mg II.  We adjust the value of C2826\prim\ by 5\% to show that it better represents the excess NUV continuum intensity at $\lambda \lesssim 2823$ \AA.}   \label{fig:sji_spec2}
\end{figure}

The full spectral readout employed for these data allow a unique and comprehensive identification of flare emission lines in the IRIS NUV wavelength range. There are many lines of Fe II in the flare spectrum in Figure \ref{fig:spectrum},
as well as for other species, and multiple lines from the same species provide powerful capability to disambiguate blends.
A number of line identifications in the flaring NUV spectrum are shown in Figures \ref{fig:sji_spec} and \ref{fig:sji_spec2}, and a full list of identifications and intensities ($<I_{\lambda}>$) at the peak of UFB-3 are given in Appendix \ref{sec:lineidsapp}.
While the strongest lines are those of Mg II, there are numerous strong lines of Fe II, Cr II, Fe I, and other species,
for which
ratios of intensity among emission lines (e.g., \feii/$\lambda$2832.39) can also 
provide valuable constraints on models.  

Another valuable capability provided by the full spectral range of these data is to characterize the contributions to SJI 2832 images in detail.
The spectral range from Figure \ref{fig:spectrum} at the peak of UFB-3 is enlarged in Figure \ref{fig:sji_spec} to show the line and continuum contributions.
The brightest emission lines in the SJI 2832 bandpass are: 
Fe II $\lambda$2826.58, $\lambda$2826.86, $\lambda$2828.26, $\lambda$2828.73, $\lambda$2829.46, $\lambda$2831.75, and $\lambda$2832.39; He I $\lambda$2829.91 (see Section \ref{sec:heliumI}); Cr II $\lambda$2831.30 and $\lambda$2833.29; Fe I $\lambda$2826.39; and Ti II $\lambda$2833.015.
Accurate
SJI 2832 predictions from RHD models of flares need to include the contribution from these lines in addition to the continuum radiation.
We calculate the filter-weighted specific intensity \citep{Sirianni2005} from the spectrum with the SJI effective area curve.  This calculation indicates that $\sim60$\% of the flaring SJI 2832 count rate is due to the NUV flare continuum radiation, which is consistent with the findings of \citet{Kleint2017} for these data.    
In Figures \ref{fig:sji_spec} and \ref{fig:sji_spec2}, we show the estimated C2826\prim\ continuum level lowered by 5\%, which seems to slightly better represent the continuum intensity between emission lines at $\lambda \lesssim 2823$ \AA.

A number of lines in the UFB-3 peak excess intensity spectrum are identified and discussed further in Appendix B. Briefly,
rest wavelengths for Fe II, Cr II, and Fe I are adopted primarily from \cite{NaveFe},  and \cite{NaveCr}, and \cite{nave_fei}, respectively, with additional data on wavelengths and $A$-values obtained from the NIST~\citep{NIST_ASD}, \cite{Kurucz2018}, and R. L. Kelly (https://www.cfa.harvard.edu/ampcgi/kelly.pl) databases and references therein. Identifications were verified by: (1) comparing line positions and profiles to those from the same species, and (2) comparing intensities from LTE calculations at different temperatures to the observed intensity ratios for each species, with opacity considered as discussed below. In Appendix B, we give measured wavelengths in this flare spectrum using 30\% bisectors.  Most bisector values for Fe II, Cr II, and Mg II are systematically redshifted by an amount greater than or approximately equal to the \feii\ line discussed in Section \ref{sec:lines}.

Nineteen Fe II line profiles, comprised by 24 of the 31 Fe II lines listed in Appendix B, are shown for the UFB-3 peak in Figure~\ref{fig:profiles}.
These are reasonably strong, isolated features with either no evidence of significant blends from other species, or minor blends that can be accounted for, since multiple other lines are observed from those species to constrain any impact the blends would have. 
This provides many ratios between Fe II line intensities that can be used to constrain model atmospheres. Compared to the ratios in the flare, the LTE intensity ratios are systematically over-predicted for the strong Fe II lines.  We speculate that optical depth in the brighter Fe II prevents a relatively larger amount of emission from escaping from the stationary flare layers \citep[below the chromospheric condensation; see ][]{Kowalski2017A} in these lines. This speculation is supported by the fact that for the two brightest Fe II lines, $\lambda$2784.512 and $\lambda$2832.394, the rest component is less pronounced relative to the red wing than it is for the weaker Fe II lines.  These two Fe II lines also exhibit the largest 30\% bisectors (7-8 km s$^{-1}$) among all Fe II line bisectors ($-1$ to $+8$ km s$^{-1}$; see Appendix B) that are reliable measured.  
This evidence for optical depth attenuation from the Fe II line ratios and profiles is consistent with the evidence from the continuum-to-line ratio (Appendix A, Section \ref{sec:c2l}, Paper II).  Detailed examination of this hypothesis can be done by simulating all the observed Fe II lines and lines of other species as discussed further in Appendix B. 

\begin{figure}[h!]
\centering
\includegraphics[scale=0.95]{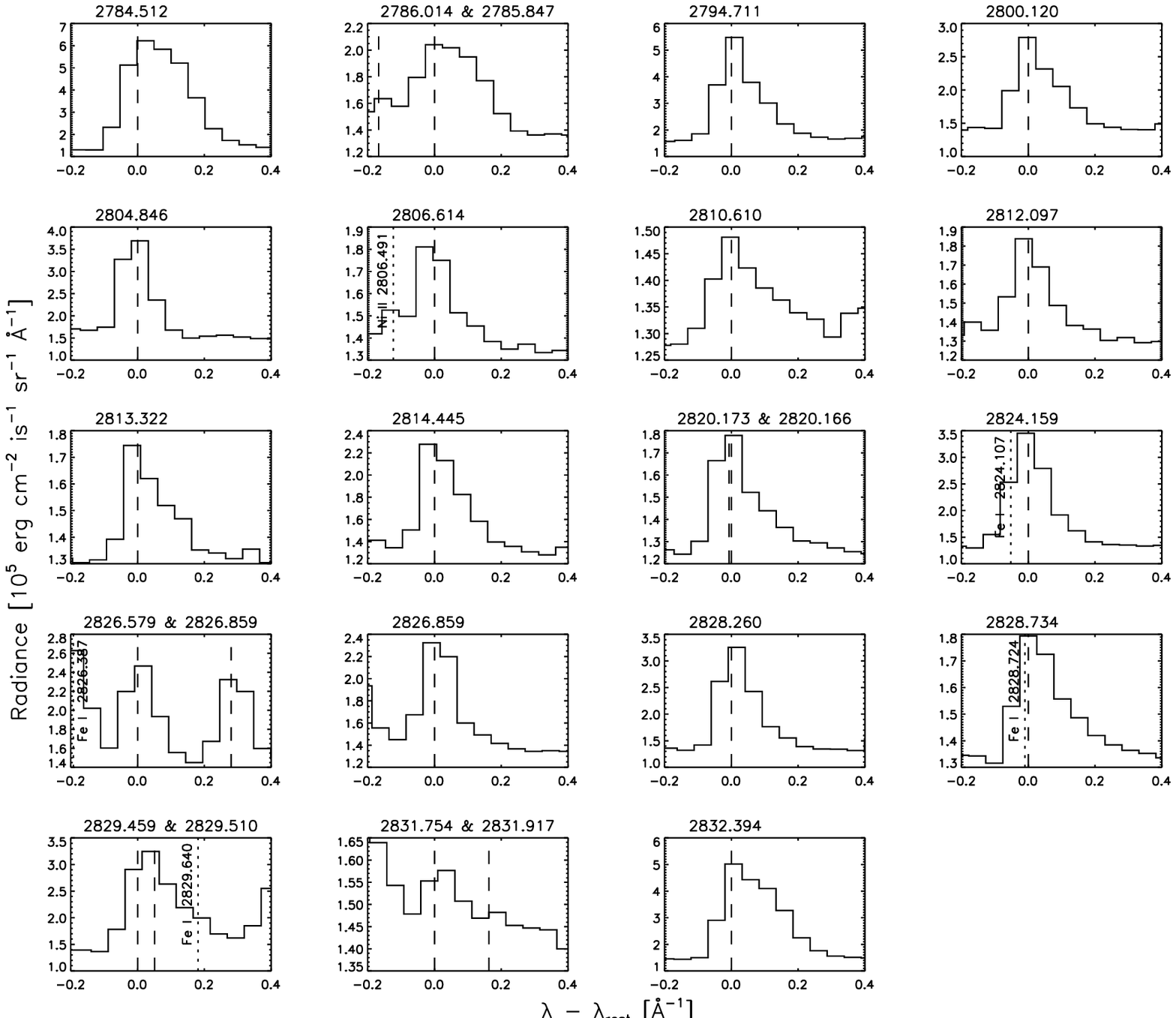}
\caption{Line profiles in the NUV spectrum of UFB-3 for all definitively identified Fe II lines.
All wavelength scales are from -0.2 to 0.4 \AA\ from the rest wavelength, where the stronger Fe II line is used when two are present.  Rest wavelengths of Fe II lines and other blends are indicated by dashed and dotted lines, respectively.  Many of the Fe II profiles exhibit a red-wing asymmetry with more spectral bins  with an intensity greater than or equal to the line half-maximum on the red side than on the blue side .} \label{fig:profiles}
\end{figure}

\subsection{A New Flare Emission Line:  He I $\lambda$2829.91} \label{sec:heliumI}
\begin{figure}[h!]
\centering
\includegraphics[scale=0.4]{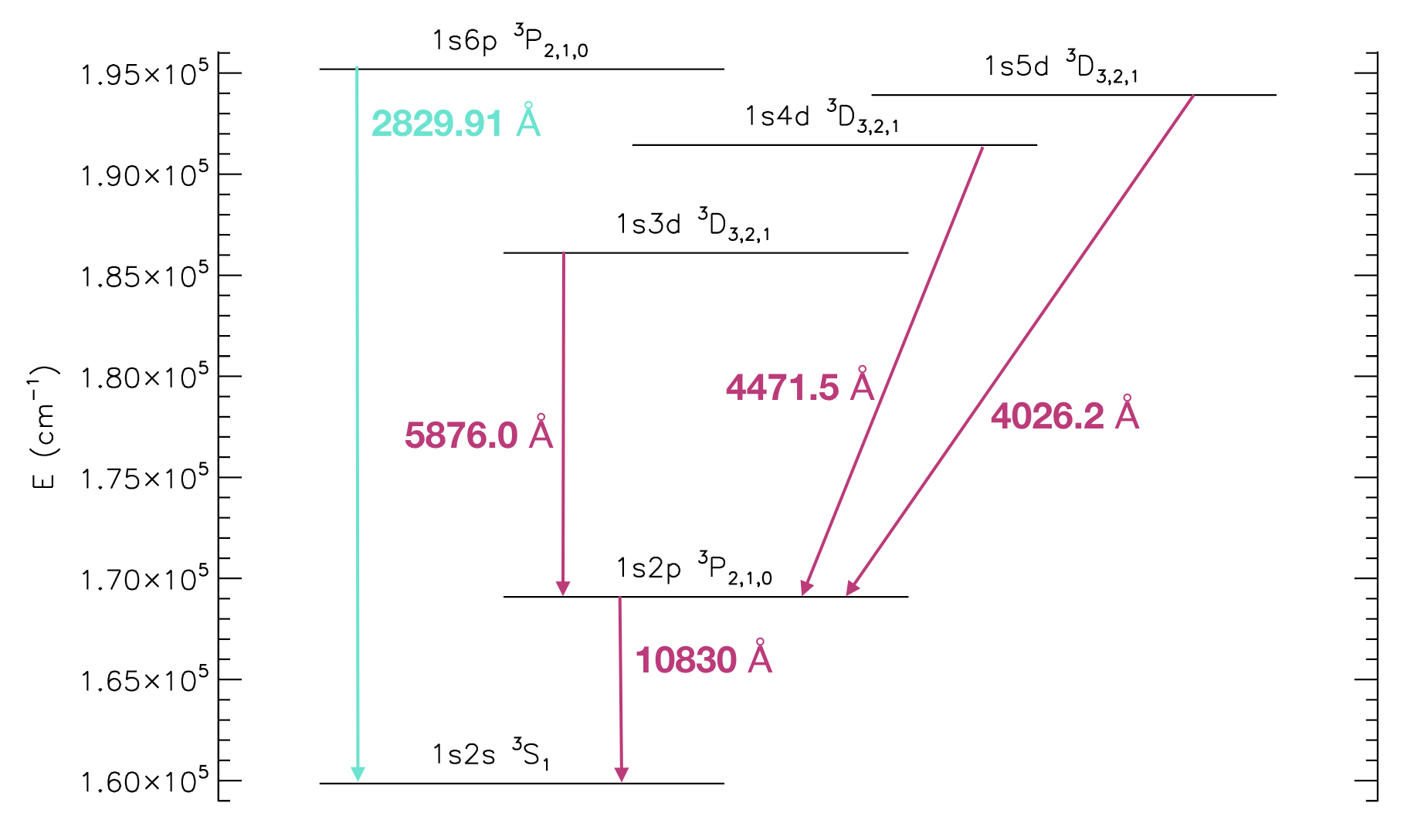}
\caption{Partial term diagram showing several important flare lines in the triplet system of He I, including the $\lambda2829.91$ multiplet in the IRIS SJI 2832 bandpass.  For reference, the ionization energy of He I is 198,310.6664 cm$^{-1}$ \citep{Kandula}. }     \label{fig:helium}
\end{figure}

We noticed an emission line in the SJI 2832 bandpass that is broader than other lines in this wavelength range.  This ``line'' is the closely-spaced He I $\lambda$2829.91 \AA\ multiplet, which has been previously identified in the laboratory  
\citep{NIST_ASD,Drake2006} and in absorption in the spectra of hot (O- and B-type) stars 
\citep{heia, heib}. This transition results from a high-lying upper level in He I compared to several other important He I flare emission lines (Figure \ref{fig:helium}).
The peak of the profile in the UFB-3 spectrum is redshifted, as for the Mg II lines.  The He I profile has comparable broadening to Mg II $\lambda$2791.6, but it exhibits a larger 30\% bisector velocity ($+11$ km s$^{-1}$) than Mg II $\lambda$2791.6 ($+8$ km s$^{-1}$) and \feii\ ($+4$ km s$^{-1}$).  The width of the profile is much larger than the instrumental resolution \citep{DePontieu2014}, which is readily apparent by comparing to the narrower Fe I and Fe II lines.  Adding the instrumental and thermal widths in quadrature, we find that the width of He I $\lambda$2829.91 equates to a thermal broadening corresponding to $T \sim 75,000$ K, which is much larger than expected for a neutral helium line.  The large FWHM (0.28 \AA) of the He I 2829.91 \AA\ line and the widths of the Mg II lines are likely due to a nonthermal broadening mechanism, such as a large optical depth and/or a spectrally unresolved redshifted component.  The larger bisector velocity than Mg II 2791.6 \AA\ and \feii\ could also be due to a larger optical depth in a chromospheric condensation:  in this case, the redshifted component to the emergent intensity would be relatively brighter compared to the component of the intensity originating from the stationary flare layers below.  There may also be a different brightness evolution in He I due to the variation in temperature sensitivity among different species formed in a condensation cooling from high to low temperature.  Ostensibly, He I lines probe higher temperatures than Fe II, Mg II, and the NUV continuum radiation but detailed modeling is necessary to account for non-equilibrium ionization of helium \citep{Allred2005, Golding2014, Allred2015, Simoes2016}.  Modeling  of this line will be included in Paper II.

\vspace{20mm}
\section{Summary \& Discussion} \label{sec:discussion}
In this paper, we present a comprehensive analysis of umbral flare ribbon spectra in the NUV.  These observations were obtained with a custom observing mode and high-time resolution, thus allowing us to characterize the flare continuum enhancements and constrain the time-evolution of heating in a single flaring location in the Sun.  We present a new spectral ratio measurement, a continuum-to-line ratio, that can be obtained with NUV solar flare spectra with very limited wavelength coverage or large uncertainties in the intensity calibration. Because the \feii\ line and the NUV continuum radiation originate from (roughly) similar temperatures and exhibit (relatively) low optical depths in current RHD flare models \citep[e.g.,][]{Kowalski2017A}, the NUV continuum-to-line ratio (C2826\prim/\feii) indicates the relative amounts of heating to $T\sim10,000$ K at high column mass (log $m$/[g cm$^{-2}$] $ \sim -2$) compared at moderate-to-low column mass (log $m$/[g cm$^{-2}$] $ \sim -3$). 
This interpretation will be described in detail in Paper II with new RHD models.  Further, the ratio is generally independent of the spatial resolution assuming that the \feii\ line and the C2826\prim\ continuum radiation are produced at the same locations and do not exhibit different spatial structures below the instrumental resolution.  Thus, the ratio facilitates a robust comparison of 1D models to observations that are spatially unresolved \citep[as in dMe flares using the ratio of H$\gamma$ line flux to the blue 4170 \AA\ continuum;][]{Kowalski2013,Silverberg2016,Kowalski2018}. or have low spatial resolution, in which case the value of the continuum intensity is not accurately inferred due to spatial smoothing.

In the umbral flare brightening spectra of the 2014-Oct-25 X1 flare, C2826\prim/\feii$ =7-8$, which is a factor of seven to eight larger than in the HXR impulsive phase spectra of the 2014-Mar-29 X1 flare:  in the ``BFP2'' flare spectrum of the 2014-Mar-29 flare \citep{Kowalski2017A}, the C2826\prim/\feii\ ratio is 1.0 but the value of C2826\prim\ is nearly a factor of 20 larger than at the peak of UFB-3. A 5F11 RHD simulation was used to model the red-wing asymmetry and bright continuum radiation in the 2014-Mar-29 flare in \citet{Kowalski2017A}.  This model predicts a continuum-to-line ratio of 1.1 (with a microturbulence parameter) and 1.8 (without a microturbulence parameter) and thus does not explain the large values in the UFBs in the 2014-Oct-25 flare.  In the 2014-Mar-29 flare, the \feii\ line exhibits a very bright and broad red wing component (see Figure \ref{fig:spectrum}(b)).  A lack of a bright, spectrally resolved red wing component in the 2014-Oct-25 flare contributes to a larger C2826\prim/\feii, but this is not nearly enough to explain the large values.   These shortcomings of the 5F11 RHD model  motivate using the Fermi/GBM data to model the nonthermal electron parameters in the 2014-Oct-25 X1 flare.  The Fermi X-ray data (Section \ref{sec:fermigbm}) suggest a similar energy flux as for the 2014-Mar-29 flare \citep{Kleint2016} but a much steeper power law index ($\delta=8-9$ compared to 4).

 The observed range of red-wing properties in X-class flares suggest that the SJI 2832 contributions from emission lines may be significantly different in other flares that produce a spectrally resolved, bright red-wing asymmetry in the Fe II, Cr II, and Helium I lines, which all contribute to the SJI 2832 bandpass.  In addition to providing detailed information on the contributions to the IRIS SJI 2832 during flares, we used the full readout data to characterize a He I flare line in the NUV. The helium emission will be used to provide constraints on the heating at higher temperatures than where Fe II and NUV continuum form, and it will help understand the origin of the non-thermal broadening in NUV flare lines.  To our knowledge, this is the first report of this He I line in a solar flare.

\section{Conclusions} \label{sec:conclusions}
We detect bona-fide NUV continuum radiation in IRIS flare spectra with a contrast of 1000\% in a solar umbra.  The spectra were obtained during the fourth hard X-ray peak at $E> 35$ keV as a ribbon (that was part of a larger, three-ribbon X-class solar flare) developed into a sunspot.  The main result of this analysis is the characterization of the ratio of NUV continuum radiation to the \feii\ line-integrated intensity, obtained from the IRIS NUV spectra.  This continuum-to-line ratio is a new diagnostic of the relative heating rates at high and low column mass in the flare chromosphere because they are formed over similar temperatures (Appendix A) with moderately different optical depths \citep[see][and Paper II]{Kowalski2017A}.  The ratios vary over values of $\approx 5-8$ in the umbral flare brightenings, attaining values of $\approx 7-8$ over the peak times.   New RHD models are required to explain these large values, since previous high-beam flux models predict much lower values and highly redshifted Fe II emission line components \citep{Kowalski2017A}.  

 With the full spectral range of IRIS, we establish, for this flare, that the intensity in the narrow continuum window from $\lambda=2824.5-2825.90$ \AA\ (C2826) adequately represents the continuum level throughout the full NUV range of IRIS.  We also identified the flare emission line landscape in the IRIS NUV and IRIS SJI 2832 images and found that the He I $\lambda2829.9$ \AA\ line becomes bright and broad in solar flares.  This line is in the SJI 2832 wavelength range and 
requires full spectral readout of the IRIS NUV, as it is not included in standard IRIS line lists.  Full spectral readout of the IRIS NUV also provides multiple lines of Fe II, Fe I, Cr II, and other species, yielding sets of profiles to constrain properties of the flaring solar atmosphere and disambiguate blends.
The emission lines exhibit red profile asymmetries, but these are much less redshifted and broad than in another well-studied X-class flare with IRIS data.

The hard X-rays from Fermi/GBM combined with an algorithm (from Graham et al., in prep) to estimate the flare area provide starting-point inputs into RHD flare models of these intriguing umbral flare brightenings, in order to determine if the IRIS NUV flare spectra can be explained by electron beam heating alone and whether significant photospheric heating is required to produce large continuum-to-line ratios.  In Paper II, we will use these two constraints (the continuum-to-line ratio and the Fe II line profile asymmetries) in addition to the constraints at higher temperatures from He I $\lambda2829.9$ to determine the relative heating in the photosphere, in the chromosphere at high column mass, and in the chromosphere at low column mass in solar flares.  

\acknowledgements
We thank an anonymous referee for comments that significantly improved the manuscript and the presentation of the results.
We gratefully acknowledge the IRIS observation planners and conversations at Dr. Paola Testa's workshop on solar microflares at the International Space Science Institute in Bern, Switzerland.  AFK and EB acknowledge support from NASA Helio GI Grant NNX17AD62G.
LF acknowledges support from the UK's Science and Technology Facilities Council under grant ST/P000533/1. 
IRIS is a NASA small explorer mission developed and operated by LMSAL with mission operations executed at NASA Ames Research center and major contributions to downlink communications funded by ESA and the Norwegian Space Center.

\appendix
\section{The formation of \feii\ and continuum radiation at $\lambda=2826$ \AA\ in LTE} \label{sec:feii}
In this appendix, we present the temperature and density sensitivities for hydrogen Balmer recombination radiation, the line intensity in \feii, and their ratios. The similar temperature sensitivities (assuming LTE) justify using the \feii\ to constrain the velocity field in the layers where the NUV continuum radiation is formed, and the ratios from optically thin uniform slabs suggest that a more sophisticated approach to the modeling (e.g., with RHD models) is necessary.

We calculate the LTE,
$\lambda=2826$ \AA\ hydrogen bound-free continuum emissivity and the \feii\ line-integrated emissivity for a range of gas densities ($\rho$) and temperatures ($T$).   We use the standard equations from \citet{Rutten2003}, \citet{Aller1963}, and \citet{Mihalas1978} (Eq. 7-4) for the spontaneous thermal, LTE line and continuum emissivities.  For hydrogen ($Z=1$) recombination to a given principle quantum number $n$, the LTE continuum emissivity, $j_{\lambda}$, follows from $\alpha(\lambda) B_{\lambda}$ (where $\alpha$ is the opacity corrected for stimulated emission) and varies as $\lambda^{-2} e^{-\frac{hc}{\lambda kT}} g_{\rm{bf}}(\lambda)$ \citep[see also Eq. 3 of][]{Kowalski2015}.  Including the $n$ dependence, the continuum emissivity reduces to the following:

\begin{equation} \label{eq:bf}
    j_{\lambda,\rm{b-f}} = \frac{6.48\times10^{-14}}{4\pi\lambda^2} \frac{n_e n_p}{T_e^{1.5} n^3} exp[ \frac{1.58\times10^5}{n^2 T_e} - \frac{1.44\times10^8}{\lambda T_e}] g_{\rm{bf}}(\lambda)
\end{equation}

\noindent in units of [erg s$^{-1}$ cm$^{-3}$ sr$^{-1}$ \AA$^{-1}$], and $[\lambda]=$\AA.  The values of the hydrogen bound-free gaunt factor ($g_{\rm{bf}}(\lambda)$) are taken from \citet{Seaton1960}.  We  calculate the recombination to $n=2$; recombination to $n=3$ and free-free emissivity contribute a moderate amount (20\%) to the continuum emissivity only at the highest temperatures that we consider ($T=22,000$ K).  At the lowest temperatures, H$^{-}$ recombination may contribute at the highest densities (but is not included here).

For the spontaneous thermal line emissivity, we use Eq. 2.69 of \citet{Rutten2003} with $n_{\rm{upper}}$ calculated from LTE using the partition function of \citet{Helenka1984} and the ionization potential lowered by 0.1 eV for Fe I and Fe II and 0.25 eV for Fe III.  We integrate over the wavelength of the line (thus, giving Eq. 2.70 of \citet{Rutten2003}).

\begin{figure}[h!]
\centering
\includegraphics[scale=0.5]{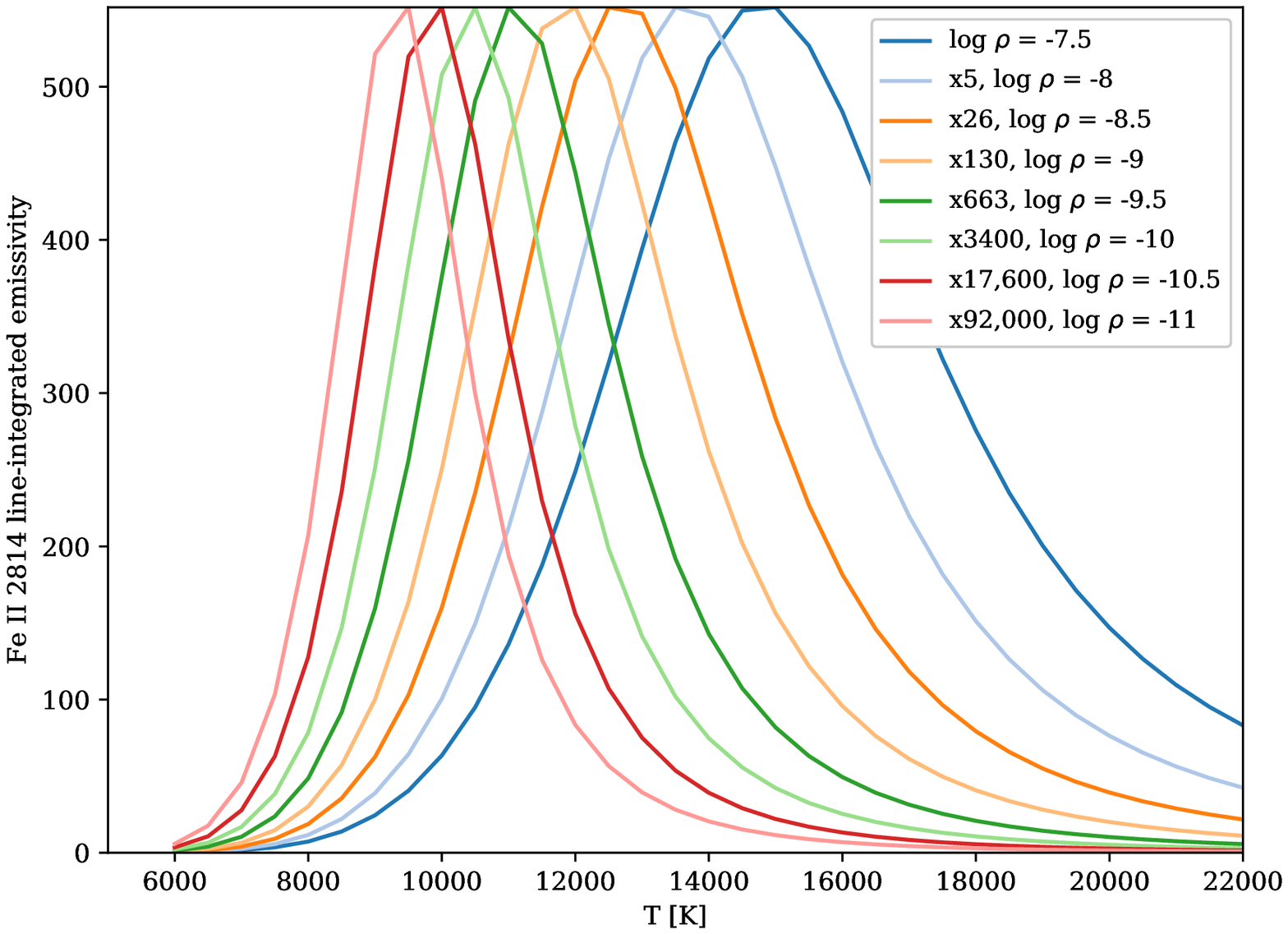}
\includegraphics[scale=0.5]{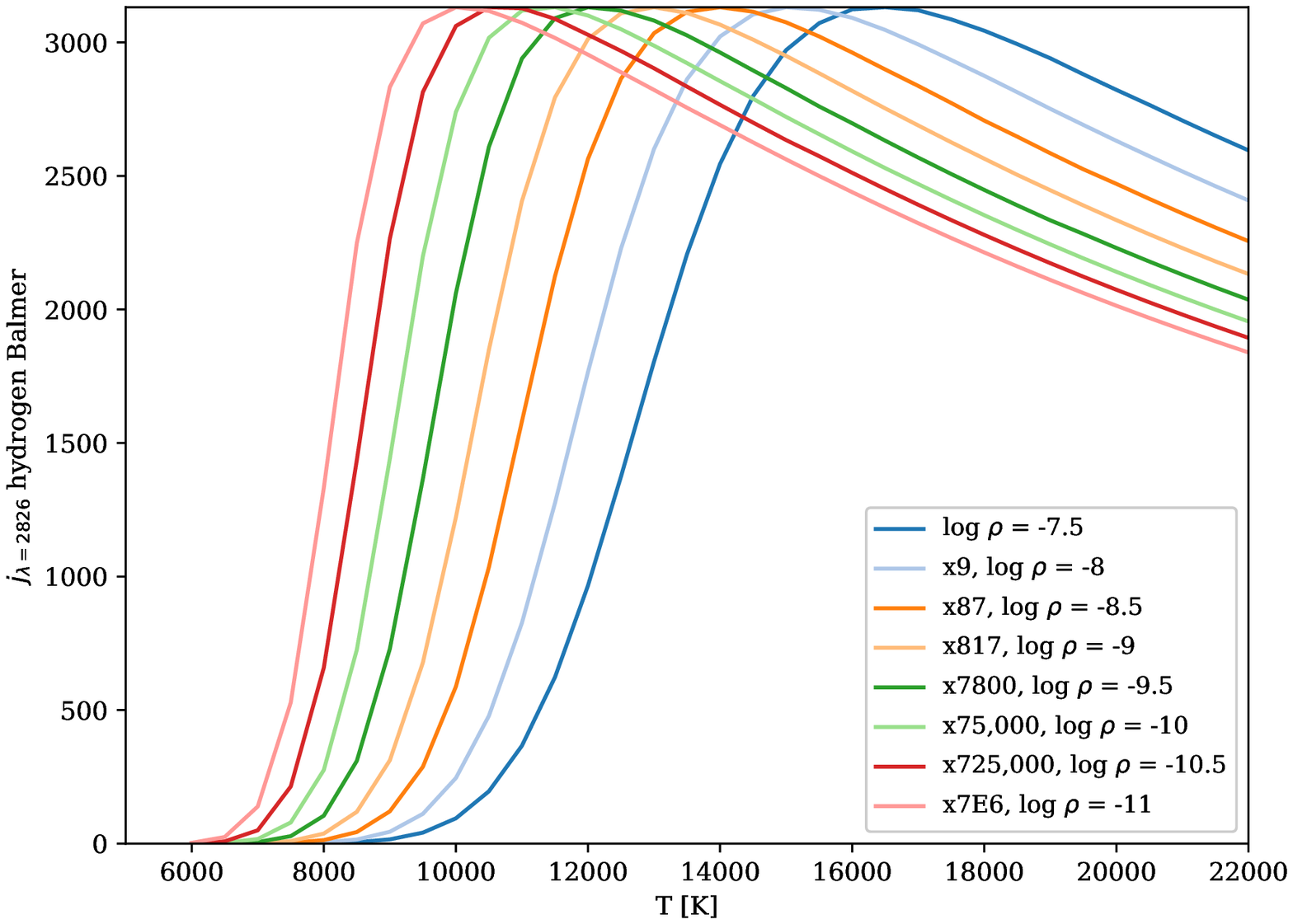}
\caption{(Top) Thermal, LTE \feii\ emissivity as a function of $\rho$ [g cm$^{-3}$] and $T$ [K].  (Bottom) Balmer continuum emissivity (recombination to $n=2$) as a function of $\rho$ and $T$, given by Equation \ref{eq:bf}. The free-free continuum emissivity and recombination to $n=3$ are not included here; they each contribute only $\sim10$\% to the total emissivity at the highest temperatures over this temperature range.  The legends indicate the scaling used to multiply each emissivity curve to the y-axis range for the log $\rho=-7.5$ curves. }   \label{fig:emiss}
\end{figure}

\begin{figure}[h!]
\centering
\includegraphics[scale=0.65]{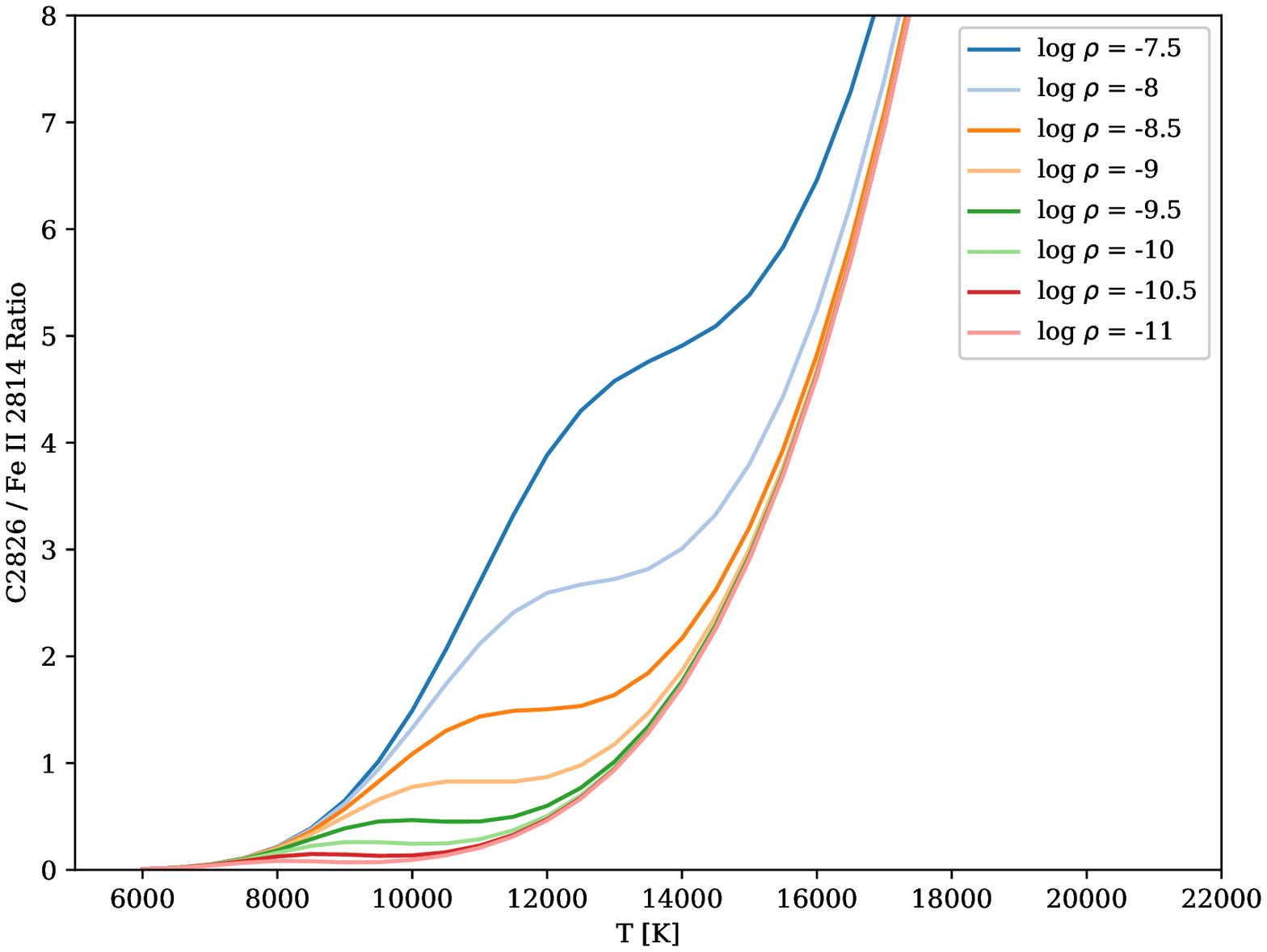}
\caption{ The C2826/\feii\ emissivity (also, optically thin LTE emergent intensity) ratios as a function of $\rho$ and $T$.   In RHD model atmospheres, the ratios of the emergent intensity differ from these values due to non-uniform $\rho$ and $T$ as a function of height and time, possible non-equilibrium effects (e.g., non-LTE ionization of \feii), and differences in the optical depth between the emergent C2826\prim\ continuum radiation and the \feii\ emission line. These slab models do not reasonably explain the observed range of ratios ($7-8$; Figure \ref{fig:lc}) in the umbral flare brightenings. }   \label{fig:ratios}
\end{figure}

The emissivity grids are shown in Figure \ref{fig:emiss}.  For the same density, the bound-free emissivity peaks at slightly higher ($\Delta T \sim 1000$ K) temperature and has a relatively brighter tail at higher temperatures.  
However, the peaks in the continuum emissivity curves are rather broad: 90\% of the maximum for each $\rho$ value falls within the peak of the \feii\ emissivity, and at 50\%, the emissivity of \feii\ spans the temperatures of $8000 - 18,000$ K.  

 In Figure \ref{fig:ratios}, we show the ratios of the emissivities.  This figure also gives the emergent intensity ratio from isothermal, isobaric optically thin, slab models in LTE.  The ratios are typically much less than unity except at very high temperatures at $T>$ 16,000 K.   In RHD models \citep{Kowalski2017A}, the \feii\ line and NUV continuum radiation are formed at significantly lower temperatures than $T = 16,000$ K because most of the flare chromospheric mass is at lower temperatures.  For example, about 95\% of the emergent NUV continuum intensity originates from $T\le18,000$ K in the evolved chromospheric condensation in the 5F11 electron beam-heated model atmosphere in \citet{Kowalski2017A}.  Therefore, optically thin slab predictions over a reasonable temperature range do not explain the large continuum-to-line ratios in the umbral flare brightenings.  As we will show in Paper II, large optical depths from significant heating at high column mass can produce consistent continuum-to-line ratios.

\section{Flare Line Identifications in the IRIS NUV} \label{sec:lineidsapp}

\begin{figure}[h!]
\centering
\includegraphics[scale=0.95]{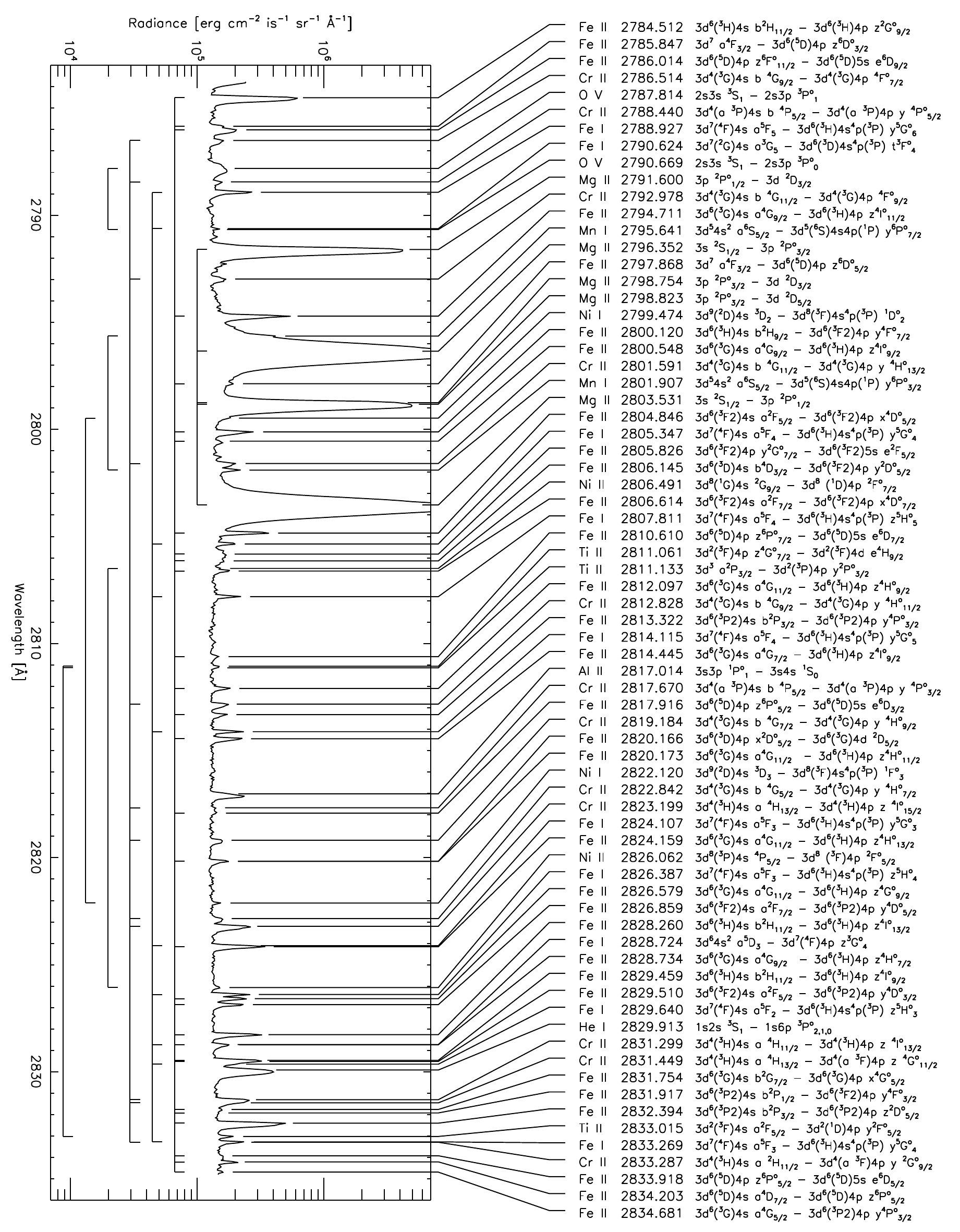}
\caption{The IRIS NUV excess spectrum at the peak of UFB-3, showing all the lines identified in Table B1.  The brackets on the left indicate multiple observable lines from the same species.}   \label{fig:ids}
\end{figure}

A number of lines in the UFB-3 peak excess intensity spectrum are identified in Table B1 and shown in Figure~\ref{fig:ids}. The line-integrated, continuum-subtracted line intensities were estimated by fitting the local continuum using a first-order polynomial and subtracting across the emission line(s). The 30\% bisectors are calculated from the intensity after subtracting a local continuum (without subtracting a pre-flare spectrum) to indicate the measured wavelengths.
In the last two rows, the excess continuum intensity, C2826\prim, and the IRIS NUV bandpass-integrated brightness are also given. Rest wavelengths for Fe II, Cr II, and Fe I are adopted primarily from \cite{NaveFe}, \cite{NaveCr}, and \cite{nave_fei}, respectively, and additional data on wavelengths and A-values were obtained from the NIST~\citep{NIST_ASD}, \cite{Kurucz2018}, and R. L. Kelly (https://www.cfa.harvard.edu/ampcgi/kelly.pl)
 databases and references therein. Most observed species have more than one line in the IRIS NUV, and many have lines from the same multiplet. Consistency for each species was established by comparing line positions and profiles to those from the same species, and by comparing the observed intensity ratios for each species to those from Boltzmann-distributed populations for a range of temperatures of formation ($e.g.,$ 7,000 to 22,000 K for Fe II; see Fig.~\ref{fig:emiss}), that is, optically thin LTE calculations, with opacity considered in the comparisons as discussed below.

Of the 31 Fe II lines listed in the UFB-3 peak spectrum, the majority are reasonably strong, isolated features with no evidence of significant blends, providing many ratios between Fe II line intensities that can be used to constrain model atmospheres. As discussed in Section~\ref{sec:lines}, compared to the ratios in the flare, 
the LTE intensity ratios are systematically over-predicted for the bright Fe II lines, and we speculate that optical depth in these brighter Fe II lines prevents a relatively larger amount of emission from escaping from the stationary flare layers below the chromospheric condensation. This speculation is supported by the fact that for the two brightest Fe II lines, $\lambda$2784.512 and $\lambda$2832.394, the rest component is less pronounced relative to the red wing than it is for the weaker Fe II lines (Figure \ref{fig:profiles}).  
 Recent 
electron impact excitation calculations by \cite{Tayal2018} are now available for 25 of the 31 observed Fe II lines, which can be used to facilitate accurate modeling of these lines to test this hypothesis.

Two observed Fe II lines are not in fact listed in the most recent work by \cite{NaveFe}: Fe II $\lambda$2820.173 and $\lambda$2828.724, although these most recent energy levels are certainly consistent with the observed wavelengths.  In the case of Fe II 2820.173, the lack of an identification in the most recent laboratory work is possibly due to blending with the Fe II 2820.166 line, which is readily excited in laboratory spectra, but comes from a higher level of excitation than would be expected for solar plasmas.  As with the other flare line identifications for UFB-3, evidence is provided by the similarity of the
Fe II $\lambda$2820.173 and $\lambda$2828.724 profiles to other Fe II lines, and of intensity levels to preliminary LTE calculations.  In this case, both these lines and Fe II $\lambda$2824.159 are from the same multiplet as well, so we are fairly confident of these identifications. There is an Fe I line at 2828.724 \AA\ to be aware of, but it is expected to be an order of magnitude weaker than the other nearby observed lines of Fe I.  The $A$-values reported by \cite{Kurucz2018} and \cite{Fuhr2006} are in reasonable agreement for all of the IRIS NUV Fe II lines that are in both databases, although with differences in some cases of up to a factor of two. $A$-values for six of the lines, including $\lambda$2828.724, are reported only by 
Kurucz, so in Table B1 the $A$-value for those lines are from Kurucz, while the balance are from Fuhr \& Wiese. 

While the most extensive set of profiles is provided by Fe II, there are quite a few Cr II lines with profiles having red wings similar to those of Fe II, which provide another set of lines to test the hypothesis above by comparison to future modeling. Multiple lines of Fe I, Ti II, Ni I, Ni II, and Mn I are observed as well. Note that, like the He I 2829.91 \AA\ line discussed in section~\ref{sec:heliumI} and Mg II h \& k, the Al II 2817.014 \AA\ line peaks to the red in UFB-3.  There is a line of Ti II in the red wing of Al II, but the intensities of the other observed lines of Ti II do not indicate that Ti II is a significant contribution to this Al II profile.  Finally, note that O V is listed not as a definitive identification, but as a possible contribution to the feature around 2788 \AA . If present, the O V 2787.814 \AA\ line appears blended with other lines, and it would be accompanied by the weaker line from the same multiplet at 2790.669 \AA , for which there is a blend with a weak Fe I line. 

Interestingly, there is a line from a doubly-excited state of He I that has been observed  at 2819.2$\pm0.3$ \AA\ using beam-foil spectroscopy \citep{Berry}, which could conceivably contribute to the observed feature at that wavelength in the UFB-3 flare spectrum. In solar plasmas, such doubly-excited states would be populated through electron capture by non-thermal (accelerated) helium ions upon hitting a thick target, but not in thermal plasmas.  Unfortunately, while there are stronger such lines outside the IRIS NUV band pass, this one has a blend with Cr II 2819.18, which is the dominant contribution to the observed feature.  It is nevertheless possible upon further evaluation of all the Cr II profiles that some evidence of a contribution from doubly-excited helium in that profile would provide a signature of accelerated helium ion beams. 

In summary, a number of IRIS NUV lines are identified in this appendix that will be useful to model and compare to the data in order to constrain the physical structure of the flaring atmosphere observed by the IRIS spectrograph.

\startlongtable
\begin{deluxetable}{lcccclrrl}   
\tablewidth{8.5in}
\tabletypesize{\scriptsize}
\tablecaption{Flare line IDs in the IRIS NUV } \label{table:lineids}
\tablehead{\colhead{Intensity} & \colhead{Species$^*$} & \colhead{$\lambda_{\rm{rest}}$} & \colhead{Bis$\dagger$} & \colhead{$J_l-J_u$} &\colhead{Multiplet}
&\colhead{$E_l$}
&\colhead{$E_u$}
&\colhead{A-value} \\
\colhead{[$10^3{\rm erg}\over{\rm{cm}^2\rm{s\ sr}}$]} & \colhead{} & \colhead{[\AA]} & \colhead{[km s$^{-1}$]} & \colhead{} & \colhead{} & \colhead{[cm$^{-1}$]} & \colhead{[cm$^{-1}$]} & \colhead{[s$^{-1}]$} }
\startdata
 112 & Fe II &  2784.512 & 7.5 & $11/2-9/2$ & $3d^6(^3$H$)4s$ b$^2$H $-$ $3d^6(^3$H$)4p$ z$^2$G$^o$  &  26170.181 &  62083.118 & 1.06e$+$08 \\
   4 & (Fe II) &  2785.847 & \nodata & $3/2-3/2$ & $3d^7$ a$^4$F $-$ $3d^6(^5$D$)4p$ z$^6$D$^o$  &   3117.488 &  39013.216 & 7.57e$+$03 \\
   19 & Fe II &  2786.014 & 4.4 & $11/2-9/2$ & $3d^6(^5$D$)4p$ z$^6$F$^o$ $-$ $3d^6(^5$D$)5s$ e$^6$D  &  41968.070 &  77861.650 & 1.53e$+$08 \\
   8 & Cr II &  2786.514 & 6.7 & $9/2-7/2$ & $3d^4(^3$G$)4s$ b$^4$G $-$ $3d^4(^3$G$)4p$ $^4$F$^o$  &  33618.936 &  69506.065 & 2.09e$+$08 \\
  24 & (O V)   &  2787.814 & 8.3 & $1-1$ & $2s3s$ $^3$S $-$ $2s3p$ $^3$P$^o$ &      546972.700 & 582843.100 & 1.41e$+$08 \\
   6 & Cr II &  2788.440 & 6.0 & $5/2-5/2$ & $3d^4(a^3$P$)4s$ b$^4$P $-$ $3d^4(a^3$P$)4p$ y$^4$P$^o$  &  30864.433 &  66726.782 & 1.50e$+$08 \\
  19 & Fe I  &  2788.927 & 0.7 & $5-6$ & $3d^7(^4$F$)4s$ a$^5$F $-$ $3d^6(^3$H$)4s4p(^3$P$^o)$ y$^5$G$^o$  &   6928.268 &  42784.349 & 6.30e$+$07 \\
 2 & (Fe I)  &  2790.624 & \nodata & $5-4$ & $3d^7(^2$G$)4s$ a$^3$G $-$ $3d^6(^3$D$)4s4p(^3$P$^o)$ t$^3$F$^o$  &  21715.731 &  57550.006 & 2.36e$+$07 \\
\nodata & (O V)   &  2790.669 & \nodata & $1-0$ & $2s3s$ $^3$S $-$ $2s3p$ $^3$P$^o$ &      546972.700 & 582806.400 & 1.43e$+$08 \\
 1175 & Mg II &  2791.600 & 8.2 & $1/2-3/2$ & $2p^63p$ $^2$P$^o$ $-$ $2p^63d$ $^2$D  &  35669.310 &  71491.063 & 4.08e$+$08 \\
   8 & Cr II &  2792.978 & 3.6 & $11/2-9/2$ & $3d^4(^3$G$)4s$ b$^4$G $-$ $3d^4(^3$G$)4p$ $^4$F$^o$  &  33694.142 &  69498.214 & 2.30e$+$08 \\
  53 & Fe II &  2794.711 & 3.2 & $9/2-11/2$ & $3d^6(^3$G$)4s$ a$^4$G $-$ $3d^6(^3$H$)4p$ z$^4$I$^o$  &  25805.327 &  61587.205 & 1.30e$+$07 \\
  1 & Mn I  &  2795.641 & \nodata & $5/2-7/2$ & $3d^54s^2$ a$^6$S $-$ $3d^5(^6$S$)4s4p(^1$P$^o)$ y$^6$P$^o$  &      0.000 &  35769.970 & 3.62e$+$08 \\
7036 & Mg II &  2796.352 & 9.9 & $1/2-3/2$ & $2p^63s$ $^2$S $-$ $2p^63p$ $^2$P$^o$  &      0.000 &  35760.880 & 2.68e$+$08 \\
  2 & (Fe II) &  2797.868 & \nodata & $3/2-5/2$ & $3d^7$ a$^4$F $-$ $3d^6(^5$D$)4p$ z$^6$D$^o$  &   3117.488 &  38858.970 & 2.02e$+$03 \\
 \nodata & Mg II &  2798.754 & \nodata & $3/2-3/2$ & $2p^63p$ $^2$P$^o$ $-$ $2p^63d$ $^2$D  &  35760.880 &  71491.063 & 8.09e$+$07 \\
1683 & Mg II &  2798.823 &  6.0 & $3/2-5/2$ & $2p^63p$ $^2$P$^o$ $-$ $2p^63d$ $^2$D  &  35760.880 &  71490.190 & 4.81e$+$08 \\
   1 & Ni I  &  2799.474 & \nodata & $2-2$ & $3d^9(^2$D$)4s$ $^3$D $-$ $3d^8(^3$F$)4s4p(^3$P$^o)$ $^1$D$^o$  &    879.813 &  36600.805 & 5.77e$+$06 \\
   20 & Fe II &  2800.120 & 3.8 & $9/2-7/2$ & $3d^6(^3$H$)4s$ b$^2$H $-$ $3d^6(^3$F$2)4p$ y$^4$F$^o$  &  26352.767 &  62065.528 & 1.55e$+$07 \\
 3 & (Fe II) & 2800.548 & 5.4 & $ 9/2 - 9/2 $ & $3d^6(^3$G$)4s$ a$^4$G $-$ $3d^6(^3$H$)4p$ z$^4$I$^o$  &  25805.327 &  61512.630 & 5.00e$+$05 \\
   10 & Cr II &  2801.591 & 7.3 & $11/2-13/2$ & $3d^4(^3$G$)4s$ b$^4$G $-$ $3d^4(^3$G$)4p$ y$^4$H$^o$  &  33694.142 &  69388.151 & 2.20e$+$08 \\
   7 & Mn I  &  2801.907 & -0.4 & $5/2-3/2$ & $3d^54s^2$ a$^6$S $-$ $3d^5(^6$S$)4s4p(^1$P$^o)$ y$^6$P$^o$  &      0.000 &  35689.980 & 3.69e$+$08 \\
6169 & Mg II &  2803.531 & 10.8 & $1/2-1/2$ & $2p^63s$ $^2$S $-$ $2p^63p$ $^2$P$^o$  &      0.000 &  35669.310 & 2.62e$+$08 \\
  23 & Fe II &  2804.846 &  -0.4 & $5/2-5/2$ & $3d^6(^3$F$2)4s$ a$^2$F $-$ $3d^6(^3$F$2)4p$ x$^4$D$^o$  &  27620.403 &  63272.981 & 1.60e$+$06 \\
   7 & Fe I  &  2805.347 & -0.5 & $4-4$ & $3d^7(^4$F$)4s$ a$^5$F $-$ $3d^6(^3$H$)4s4p(^3$P$^o)$ y$^5$G$^o$  &   7376.764 &  43022.982 & 1.05e$+$07 \\
 1 & (Fe II) &  2805.826 & \nodata & $ 7/2 - 5/2 $ & $3d^6(^3$F$2)4p$ y$^2$G$^o$ $-$ $3d^6(^3$F$2)5s$ e$^2$F  &  65109.691 & 100749.825 & 1.39e$+$08 \\
 1 & (Fe II) &  2806.145 & \nodata & $ 3/2 - 5/2 $ & $3d^6(^3$D$)4s$ b$^4$D $-$ $3d^6(^3$F$2)4p$ y$^2$D$^o$  &  31364.455 &  67000.530 & 2.50e$+$06 \\
  1 & Ni II &  2806.491 & \nodata & $9/2-7/2$ & $3d^8(^1$G)$4s$ $^2$G $-$ $3d^8(^1$D)$4p$ $^2$F$^o$ &  32499.530 &  68131.210 & 1.30e$+$07 \\
   5 & Fe II &  2806.614 & -0.9 & $7/2-7/2$ & $3d^6(^3$F$2)4s$ a$^2$F $-$ $3d^6(^3$F$2)4p$ x$^4$D$^o$  &  27314.918 &  62945.045 & 3.20e$+$06 \\
   9 & Fe I  &  2807.811 & 0.0 & $4-5$ & $3d^7(^4$F$)4s$ a$^5$F $-$ $3d^6(^3$H$)4s4p(^3$P$^o)$ z$^5$H$^o$  &   7376.764 &  42991.694 & 1.15e$+$07 \\
   4 & Fe II &  2810.610 & 5.9 & $7/2-7/2$ & $3d^6(^5$D$)4p$ z$^6$P$^o$ $-$ $3d^6(^5$D$)5s$ e$^6$D  &  42658.244 &  78237.709 & 3.10e$+$07 \\
  3 & Ti II &  2811.061 & 2.6 & $7/2-9/2$ & $3d^2(^3$F$)4p$ z$^4$G$^o$ $-$ $3d^2(^3$F$)4d$ e$^4$H  &  29734.540 &  65308.300 & 5.09e$+$08 \\
1 & Ti II &  2811.133 & \nodata & $3/2-3/2$ & $3d^3$ a$^2$P $-$ $3d^2(^3$P$)4p$ y$^2$P$^o$  &   9975.920 &  45548.760 & 1.24e$+$07 \\
   8 & Fe II &  2812.097 & 0.8 & $11/2-9/2$ & $3d^6(^3$G$)4s$ a$^4$G $-$ $3d^6(^3$H$)4p$ z$^4$H$^o$  &  25428.789 &  60989.444 & 1.20e$+$06 \\
   9 & Cr II &  2812.828 & 7.9 & $9/2-11/2$ & $3d^4(^3$G$)4s$ b$^4$G $-$ $3d^4(^3$G$)4p$ y$^4$H$^o$  &  33618.936 &  69170.353 & 2.05e$+$08 \\
   6 & Fe II &  2813.322 & 4.8 & $3/2-3/2$ & $3d^6(^3$P$2)4s$ b$^2$P $-$ $3d^6(^3$P$2)4p$ y$^4$P$^o$  &  25787.582 &  61332.753 & 2.90e$+$06 \\
   11 & Fe I  &  2814.115 & 0.1 & $4-5$ & $3d^7(^4$F$)4s$ a$^5$F $-$ $3d^6(^3$H$)4s4p(^3$P$^o)$ y$^5$G$^o$  &   7376.764 &  42911.914 & 3.42e$+$07 \\
   15 & Fe II &  2814.445 & 3.7 & $7/2-9/2$ & $3d^6(^3$G$)4s$ a$^4$G $-$ $3d^6(^3$H$)4p$ z$^4$I$^o$  &  25981.645 &  61512.630 & 3.40e$+$06 \\
  22 & Al II &  2817.014 & 10.9 & $1-0$ & $3s3p$ $^1$P$^o$ $-$ $3s4s$ $^1$S  &  59852.020 &  95350.600 & 3.93e$+$08 \\
 1 & Cr II &  2817.670 & \nodata & $5/2-3/2$ & $3d^4(a^3$P$)4s$ b$^4$P $-$ $3d^4(a^3$P$)4p$ y$^4$P$^o$  &  30864.433 &  66354.757 & 1.04e$+$08 \\
 1 & (Fe II) &  2817.916 & \nodata & $ 5/2 - 3/2 $ & $3d^6(^5$D$)4p$ z$^6$P$^o$ $-$ $3d^6(^5$D$)5s$ e$^6$D  &  43238.607 &  78725.822 & 3.40e$+$07 \\
   7 & Cr II &  2819.184 & 7.4 & $7/2-9/2$ & $3d^4(^3$G$)4s$ b$^4$G $-$ $3d^4(^3$G$)4p$ y$^4$H$^o$  &  33521.090 &  68992.347 & 2.21e$+$08 \\
\nodata & Fe II &  2820.166 & \nodata & $5/2-5/2$ & $3d^6(^3$D$)4p$ x$^2$D$^o$ $-$ $3d^6(^3$G$)4d$ $^2$D  &  74606.864 & 110065.766 & 8.25e$+$05 \\
  8 & Fe II &  2820.173 & 2.7 & $11/2-11/2$ & $3d^6(^3$G$)4s$ a$^4$G $-$ $3d^6(^3$H$)4p$ z$^4$H$^o$  &  25428.784 &  60887.598 & 1.00e$+$06 \\
    2 & Ni I  &  2822.120 & 0.6 & $3-3$ & $3d^9(^2$D$)4s$ $^3$D $-$ $3d^8(^3$F$)4s4p(^3$P$^o)$ $^1$F$^o$  &    204.786 &  35639.148 & 4.87e$+$06 \\
   6 & Cr II &  2822.842 & 7.2 & $5/2-7/2$ & $3d^4(^3$G$)4s$ b$^4$G $-$ $3d^4(^3$G$)4p$ y$^4$H$^o$  &  33417.981 &  68843.273 & 2.29e$+$08 \\
  28 & Cr II &  2823.199 & 7.6 & $13/2-15/2$ & $3d^4(^3$H$)4s$ a$^4$H $-$ $3d^4(^3$H$)4p$ z$^4$I$^o$  &  30391.831 &  65812.649 & 2.28e$+$08 \\
 \nodata & Fe I  &  2824.107 & \nodata & $3-3$ & $3d^7(^4$F$)4s$ a$^5$F $-$ $3d^6(^3$H$)4s4p(^3$P$^o)$ y$^5$G$^o$  &   7728.059 &  43137.484 & 1.51e$+$07 \\
 30 & Fe II &  2824.159 & 0.2 & $11/2-13/2$ & $3d^6(^3$G$)4s$ a$^4$G $-$ $3d^6(^3$H$)4p$ z$^4$H$^o$  &  25428.789 &  60837.560 & 2.10e$+$06 \\
   3 & Ni II &  2826.062 & -0.3 & $5/2-5/2$ & $3d^8(^3$P)$4s$ $^4$P $-$ $3d^8(^3$F)$4p$ $^2$F$^o$ &  23108.280 &  58493.210 & 2.65e$+$06 \\
   14 & Fe I  &  2826.387 & 1.3 & $3-4$ & $3d^7(^4$F$)4s$ a$^5$F $-$ $3d^6(^3$H$)4s4p(^3$P$^o)$ z$^5$H$^o$  &   7728.059 &  43108.914 & 1.32e$+$07 \\
 18 & Fe II &  2826.579 & 0.5 & $11/2-9/2$ & $3d^6(^3$G$)4s$ a$^4$G $-$ $3d^6(^3$H$)4p$ z$^4$G$^o$  &  25428.789 &  60807.239 & 1.40e$+$06 \\
   16 & Fe II &  2826.859 & 1.6 & $7/2-5/2$ & $3d^6(^3$F$2)4s$ a$^2$F $-$ $3d^6(^3$P$2)4p$ y$^4$D$^o$  &  27314.918 &  62689.874 & 4.50e$+$06 \\
  25 & Fe II &  2828.260 & 2.3 & $11/2-13/2$ & $3d^6(^3$H$)4s$ b$^2$H $-$ $3d^6(^3$H$)4p$ z$^4$I$^o$  &  26170.181 &  61527.610 & 2.40e$+$06 \\
\nodata & Fe I  &  2828.724 & \nodata & $3-4$ & $3d^64s^2$ a$^5$D $-$ $3d^7(^4$F$)4p$ z$^3$G$^o$  &    415.933 &  35767.562 & 1.48e$+$05 \\
 7 & Fe II &  2828.734 & 5.2 & $9/2-7/2$ & $3d^6(^3$G$)4s$ a$^4$G $-$ $3d^6(^3$H$)4p$ z$^4$H$^o$ &  25805.328 &  61156.835 & 1.51e$+$06 \\
 36 & Fe II &  2829.459 & 7.9 & $11/2-9/2$ & $3d^6(^3$H$)4s$ b$^2$H $-$ $3d^6(^3$H$)4p$ z$^4$I$^o$  &  26170.181 &  61512.630 & 6.90e$+$06 \\
 \nodata & Fe II &  2829.510 & \nodata & $5/2-3/2$ & $3d^6(^3$F$2)4s$ a$^2$F $-$ $3d^6(^3$P$2)4p$ y$^4$D$^o$  &  27620.403 &  62962.215 & 9.00e$+$06 \\
\nodata & Fe I  &  2829.640 & \nodata & $2-3$ & $3d^7(^4$F$)4s$ a$^5$F $-$ $3d^6(^3$H$)4s4p(^3$P$^o)$ z$^5$H$^o$  &   7985.784 &  43325.961 & 1.87e$+$06 \\
\nodata   & He I  &  2829.911 & \nodata & $1-0$ & $1s2s$ $^3$S $-$ $1s6p$ $^3$P$^o$ & 159855.974 & 195192.777 & 1.94e$+$06 \\
 81  & He I  &  2829.913 & 10.9 & $1-1$ & $1s2s$ $^3$S $-$ $1s6p$ $^3$P$^o$ & 159855.974 & 195192.746 &  1.94e$+$06 \\
\nodata   & He I  &  2829.914 & \nodata & $1-2$ & $1s2s$ $^3$S $-$ $1s6p$ $^3$P$^o$ & 159855.974 & 195192.743 & 1.94e$+$06 \\
 27 & Cr II &  2831.299 & 9.3 & $11/2-13/2$ & $3d^4(^3$H$)4s$ a$^4$H $-$ $3d^4(^3$H$)4p$ z$^4$I$^o$  &  30298.468 &  65617.946 & 2.54e$+$08 \\
\nodata & Cr II &  2831.449 & \nodata & $13/2-11/2$ & $3d^4(^3$H$)4s$ a$^4$H $-$ $3d^4(a^3$F$)4p$ z$^4$G$^o$  &  30391.831 &  65709.442 & 3.71e$+$07 \\
  3 & Fe II &  2831.754 & \nodata & $7/2-5/2$ & $3d^6(^3$G$)4s$ b$^2$G $-$ $3d^6(^3$G$)4p$ x$^4$G$^o$  &  30764.474 &  66078.272 & 7.00e$+$05 \\
\nodata  & Fe II &  2831.917 & \nodata & $1/2-3/2$ & $3d^6(^3$P$2)4s$ b$^2$P $-$ $3d^6(^3$F$2)4p$ y$^4$F$^o$  &  26932.735 &  62244.515 & 5.43e$+$05 \\
  71 & Fe II &  2832.394 & 7.3 & $3/2-5/2$ & $3d^6(^3$P$2)4s$ b$^2$P $-$ $3d^6(^3$P$2)4p$ z$^2$D$^o$  &  25787.582 &  61093.406 & 7.60e$+$07 \\
   6 & Ti II &  2833.015 & 2.6 & $5/2-5/2$ & $3d^2(^3$F$)4s$ a$^2$F $-$ $3d^2(^1$D$)4p$ y$^2$F$^o$  &   4628.580 &  39926.660 & 2.46e$+$07 \\
 11 & Fe I  &  2833.269 & 3.7 & $3-4$ & $3d^7(^4$F$)4s$ a$^5$F $-$ $3d^6(^3$H$)4s4p(^3$P$^o)$ y$^5$G$^o$  &   7728.059 &  43022.982 & 2.38e$+$07 \\
\nodata & Cr II &  2833.287 & \nodata & $11/2-9/2$ & $3d^4(^3$H$)4s$ a$^2$H $-$ $3d^4(a^3$F$)4p$ y$^2$G$^o$  &  34812.926 &  70107.623 & 1.29e$+$08 \\
  1 & (Fe II) & 2833.918 & \nodata & $ 5/2 - 5/2 $ & $3d^6(^5$D$)4p$ z$^6$P$^o$ $-$ $3d^6(^5$D$)5s$ e$^6$D  &  43238.607 &  78525.442 & 4.50e$+$07 \\
   7 & (Fe II) &  2834.203 & 11.5 & $7/2-5/2$ & $3d^6(^5$D$)4s$ a$^4$D $-$ $3d^6(^5$D$)4p$ z$^6$P$^o$  &   7955.319 &  43238.607 & 4.41e$+$04 \\
  1 & (Fe II) &  2834.681 & \nodata & $5/2-3/2$ & $3d^6(^3$G$)4s$ a$^4$G $-$ $3d^6(^3$P$2)4p$ y$^4$P$^o$  &  26055.412 &  61332.753 & 1.33e$+$05 \\
   124 & C2826\prim & $2824.5 - 2825.9$ & \nodata & \nodata & \nodata & & & \\
 6200 &  NUV cont & $2785 - 2835$ & \nodata & \nodata  & \nodata & & & \\
\enddata
\tablecomments{
$^*$ Parentheses indicate identifications that are not definitive. No intensity value is given for lines that are highly blended and are not a primary contribution to the observed feature.   $^\dagger$Measured wavelengths for this spectrum are given in the \emph{Bis} column, which is the 30\% bisector of a line. 
Bisector values are not given for lines that are likely not a main contribution to the observed feature, and they are not given for very low line intensities.  For some fainter lines, a value of 1 can be estimated for the intensity.  For Fe II 2786.014, 2806.614 and  2826.579, we use a Gaussian centroid for the measured wavelength  because the 30\% bisector extends across other blended lines.  We adopt a representative centroiding uncertainty of 2.3-3.2 km s$^{-1}$ as our statistical uncertainty, and the systematic uncertainty from the centroid of the Ni I absorption line in the quiet Sun (see Section \ref{sec:irisdata}) is 1.8 km s$^{-1}$.  Though Fe II 2829.459 has the largest Fe II bisector, other lines are possibly blended with this line (see Figure \ref{fig:profiles}).   The value of C2826\prim\ is given in units of $10^3$ erg cm$^{-2}$ s$^{-1}$ sr$^{-1}$ \AA$^{-1}$.  }
\end{deluxetable}

\end{document}